\documentclass[fleqn,usenatbib,useAMS]{mnras}

\usepackage{newtxtext,newtxmath}
\usepackage[T1]{fontenc}
\usepackage{ae,aecompl}
\usepackage{caption}
\usepackage{threeparttable}
\usepackage{numprint}
\usepackage{siunitx}
\usepackage[pdftex]{graphicx}
\usepackage{subfig}
\usepackage{xcolor}

\newcommand{\orcid}[1]{\href{https://orcid.org/#1}{\includegraphics[width=10pt]{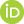}}}

\title[Eclipsing White Dwarfs in Gaia and ZTF]{Eclipsing White Dwarf Binaries in \textit{Gaia} and the Zwicky Transient Facility}

\author[P. M. Keller et al.]{
Pascal M. Keller\orcid{0000-0002-7681-8437}$^{1,2}$\thanks{E-mail: pmk46@cam.ac.uk},
Elm\'{e} Breedt\orcid{0000-0001-6180-3438}$^{2}$\thanks{E-mail: ebreedt@ast.cam.ac.uk},
Simon Hodgkin\orcid{0000-0002-5470-3962}$^{2}$,
Vasily Belokurov$^{2}$, 
James Wild\orcid{0000-0002-9967-590X}$^{3}$, 
\newauthor
Ignacio Garc\'{i}a-Soriano$^{4}$,
Jacob L. Wise\orcid{0000-0003-0733-2916}$^{3,4}$
\\
$^1$ Astrophysics Group, Cavendish Laboratory, University of Cambridge, Cambridge CB3 0HE, UK\\
$^2$ Institute of Astronomy, University of Cambridge, Madingley Road, Cambridge CB3 0HA, UK\\
$^3$ Department of Physics and Astronomy, University of Sheffield, Sheffield S3 7RH, UK\\
$^4$ Isaac Newton Group of Telescopes, Apartado de Correos 321, Santa Cruz de La Palma, E-38700, Spain
}

\date{Accepted XXX. Received YYY; in original form ZZZ}

\pubyear{2021}

\begin{document}
\label{firstpage}
\pagerange{\pageref{firstpage}--\pageref{lastpage}}

\maketitle

\begin{abstract}
  \textit{Gaia} provided the largest-ever catalogue of white dwarf stars. We use this catalogue, along with the third public data release of the Zwicky Transient Facility (ZTF), to identify new eclipsing white dwarf binaries. Our method exploits light curve statistics and the Box Least Squares algorithm to detect periodic light curve variability. The search revealed 18 new binaries, of which 17 are eclipsing. We use the position in the \textit{Gaia} H-R diagram to classify these binaries and find that the majority of these white dwarfs have main sequence companions. We identify one system as a candidate eclipsing white dwarf--brown dwarf binary and a further two as extremely low mass (ELM) white dwarf binaries. We also provide identification spectroscopy for 17 of our 18 binaries. Running our search method on mock light curves with real ZTF sampling, we estimate our efficiency of detecting objects with light curves similar to the ones of the newly discovered binaries. Many more binaries are to be found in the ZTF footprint as the data releases grow, so our survey is ongoing. 
\end{abstract}

\begin{keywords}
stars: binaries: close -- stars: binaries: eclipsing -- stars: white dwarfs
\end{keywords}

\section{Introduction}
\label{introduction}

The high precision astrometric measurements of the \textit{Gaia}\, satellite revealed evidence for stellar multiplicity across the Hertzsprung–Russell (H-R) diagram \citep{Belokurov2020}, with the binary fraction decreasing from the most massive stars towards the bottom of the main sequence and the white dwarfs. Nevertheless, binaries with white dwarf components are commonly found, although in many cases the binaries are unresolved and indirect techniques are necessary to determine the nature of the companion.
There are a variety of possible evolutionary outcomes for a pair of main sequence stars, depending on, for example, the initial mass ratio and initial separation of the component stars. This typically includes Cataclysmic Variables (CVs), AM\,CVn stars, supersoft sources, detached white dwarf main sequence binaries, white dwarf-subdwarf binaries, extremely low mass (ELM) white dwarf binaries and double white dwarfs. Among these binaries are the progenitors of Type Ia Supernova (SN\,Ia) explosions and the related .Ia supernovae, the nature of which remain among the most important unsolved problems in binary evolution.  Populations of white dwarf binaries are evolutionary connected, through complex physical processes such as common envelope evolution \citep[][and references therein]{Ivanova13,Toloza19}. Hence understanding the various populations of white dwarf binaries is an important step towards understanding the various progenitor routes towards SN\,Ia. 

Binaries which are close enough to interact will start mass transfer during their evolution and form a common envelope around both stars. This process reduces the orbital period even further and results in variety of possible binary types emerging from the common envelope. 
White dwarf--main sequence binaries  are one of the more common outcomes of this process and they play an important role in our understanding of binary evolution. As they are not accreting, they are ideal for testing various stages of the evolutionary process. For example, the angular momentum loss prescription \citep{SchreiberGaensicke03}, efficiency of the common envelope ejection \citep{Zorotovic14} and the age–metallicity relation of the solar neighbourhood \citep{RebassaMansergas21} all rely on detailed observations of these binaries. 

Recently, double white dwarfs have also been an area of particular focus, as these compact systems emit gravitational waves at mHz frequencies as they spiral together. Depending on the masses of the component stars, a white dwarf binary with a period of $\sim6-12$\,hr will merge within a Hubble time. Along with the AM\,CVn stars \citep{Ramsay18}, the most compact of these binaries will be among the first sources detected by the Laser Interferometric Space Antenna (LISA) \citep{korol2017, kupfer2018} and the wider binaries will set the gravitational wave background against which other detections will be made. 
In recent years, many of the double white dwarfs discovered contained an ELM white dwarf, following a dedicated spectroscopic survey for these objects \citep[][and earlier papers in the series]{brown2020elm}. The masses of ELM white dwarfs are too low ($M<0.3M_\odot$) that they could have formed through single star evolution -- the Universe is not old enough. Instead they are believed to form as a result of binary interaction. Observationally, the binary fraction of ELM white dwarfs is consistent with 100\% \citep{brown2020elm} although the companion is often only revealed through radial velocity variations rather than directly observable spectroscopic features. This is because the low mass of these white dwarfs imply large radii, and hence high luminosities, which often dominate the luminosity of the binaries they are part of. 

An alternative method to identify binary stars is through eclipses. Eclipsing binaries are particularly valuable, since they can yield precise parameters for the component stars, which can then be compared to models of stellar structure and binary evolution. 
The Zwicky Transient Facility (ZTF) has been particularly successful in uncovering new ultra-compact eclipsing binaries this way, notably the 6.91~min eclipsing double white dwarf ZTF\,J153932.16+502738.8 \citep{burdge2019} and 8.8\,min system ZTF\,J224342.97+524205.9 \citep{Burdge20_9min}. These are currently the shortest period eclipsing white dwarf binaries known, and they are expected to be detected by LISA in the first week and the first month of operations, respectively. Similarly,  ZTF\,J190125.42+530929.5 was confirmed as a 40.6~min eclipsing double white dwarf by \citet{coughlin2020} and \citet{Burdge20} report a further 15 variable white dwarf binaries discovered with ZTF photometry, of which nine are eclipsing. 

The majority of ZTF searches for variable white dwarfs have focused on short periods ($\lesssim$1\,hr), in order to identify ultra-compact white dwarf binaries that will be bright gravitational wave sources in the LISA band. In this work, we extend that period range by searching for binaries with periods up to 2.4\,hr. Although we use a white dwarf catalogue as starting point (Section~\ref{sec:data}), the wider period range searched is expected to reveal a variety of binary types, including white dwarfs with main sequence companions that will not necessarily be strong gravitational wave sources.

Section~\ref{sec:data} gives a brief overview of the ZTF and the \textit{Gaia} Data Release 2 white dwarf catalogue that we used in this work and Section~\ref{sec:id} describes the methods used to detect periodically variable sources. We present our results in Section~\ref{sec:results} and then discuss the 18 new binaries found in Section~\ref{sec:discussion}. We summarise in Section~\ref{sec:summary}.

\section{Observational Data}
\label{sec:data}

ZTF is a time-domain survey aimed at exploring the variable and transient universe. It uses a wide-field CCD camera mounted on the 48-inch Samuel Oschin telescope at the Palomar Observatory in California. With its 47 square-degree field of view and eight second readout time it is able to scan the entire northern sky in just two observing nights, achieving a depth of 20.5 magnitudes in the 30 second exposure time. The camera is also equipped with three custom-made filters: ZTF-$g$, ZTF-$r$ and ZTF-$i$ \citep{bellm2018, masci2019}. Since its first light in 2017 November it has produced more than $2\times 10^{9}$ light curves, taken as part of a mixture of public and private collaboration surveys. The data used in this work was taken from the third Public Data Release (ZTF DR3)\footnote{ZTF Public Data Release 3: \url{www.ztf.caltech.edu/page/dr3}}. 

In addition to ZTF, we also made use of data from the astrometric space mission \textit{Gaia} \citep[\textit{Gaia} DR2,][]{gaia2018}. The spacecraft was launched by the European Space Agency (ESA) in 2013 December. It measures the positions, parallaxes and proper motions of about $1\%$ of all stars contained in the Milky Way and reaches a depth of $G\sim20.5$ magnitudes. With the measured parallaxes and photometric data in three bands it is possible to infer the absolute magnitude and the colour of a source. Hence, sources can be located in the Hertzsprung-Russell \mbox{(H-R)} diagram, ultimately allowing a separation of white dwarfs from main sequence stars. We used the published catalogue compiled by \citet{gentile2019} to identify white dwarfs. The catalogue contains 486\,641 sources which were selected based on cuts in absolute magnitude, colour and \textit{Gaia} quality flags. Moreover, they used a set of 21\,456 spectroscopically confirmed white dwarfs from the Sloan Digital Sky Survey \citep[SDSS DR14,][]{abolfathi2018} and 5982 known contaminants in order to map a probability of being a white dwarf to each source. The final catalogue contains 260\,000 high confidence white dwarfs. 

We cross-matched the \textit{Gaia} white dwarf catalogue with ZTF DR3 in order to identify photometrically variable white dwarfs. The aperture size used in this cross-match is 1$\arcsec$. Since the catalogue is constructed to be inclusive it allows for a wider selection in the H-R diagram than just the white dwarf sequence, and we may expect to find a variety of white dwarf binaries this way. 
276\,074 (57 per cent) of the sources contained in the white dwarf catalogue have at least two ZTF epochs. This is mainly because \textit{Gaia} (and hence the white dwarf catalogue) is an all-sky survey, while ZTF only covers the sky north of $\delta>-30\degr$. 

In the next section we describe the details of the cross-match as well as our method to select objects which display variability. 

We also obtained follow-up spectroscopy of most of our binary candidates (described below) on the nights of 2021 August 4--6. We used the Intermediate Dispersion Spectrograph (IDS) on the 2.5\,m Isaac Newton Telescope (INT), located at the Roque de los Muchachos Observatory on La Palma, Spain. All spectra were obtained with an identical instrument setup, using the R300V grism centred on 6130\AA. This gives a useful spectral range of approximately 3540--8585\AA\, at a resolution of 5.5\AA. The wavelength calibration was derived from CuAr and CuNe arc lamp exposures taken after each target spectrum and adjusted using the position of the 5577\AA\, skyline. The flux calibration used an observation of the standard star Feige\,110, taken at the end of each night. We reduced the spectra using the {\sc starlink} package {\sc pamela} and applied the calibrations using {\sc molly}\footnote{Available from http://deneb.astro.warwick.ac.uk/phsaap/software/}, written by T.R.\,Marsh.
These spectra, plus one spectrum available from SDSS, provides spectroscopic identification for all but one of our targets. A log of the observations are shown in Table~\ref{tab:specobs}.

\section{Binary identification}
\label{sec:id}

This work exploits eclipses in order to detect white dwarf binaries. Because of the small size of white dwarfs, eclipses last only a short time, or last only a small fraction of the binary orbit. Therefore, in the ZTF light curves, the eclipses present as one or two points occasionally dipping below the majority of points in the light curve. For a sufficiently high inclination, one binary cycle consists of two eclipses, a deeper primary and a shallower secondary, separated by half the orbital period. The primary eclipse is generated when the brighter object is obscured by its companion and vice versa for the secondary eclipse. The secondary eclipse, however, is in some cases too shallow to be detected. A method well suited to detecting eclipses is the Box Least Squares algorithm (BLS), which was developed in order to search for periodic transits of extrasolar planets \citep{kovacs2002}. BLS is able to detect eclipses by obtaining the best fit of a periodic box with a characteristic duration and depth. In this work we used the astropy implementation of BLS \citep{astropy2013, astropy2018}.

\subsection{Period Search}
\label{sec:BLS}

\begin{figure}
	\centering
	    \includegraphics[width=.48\textwidth]{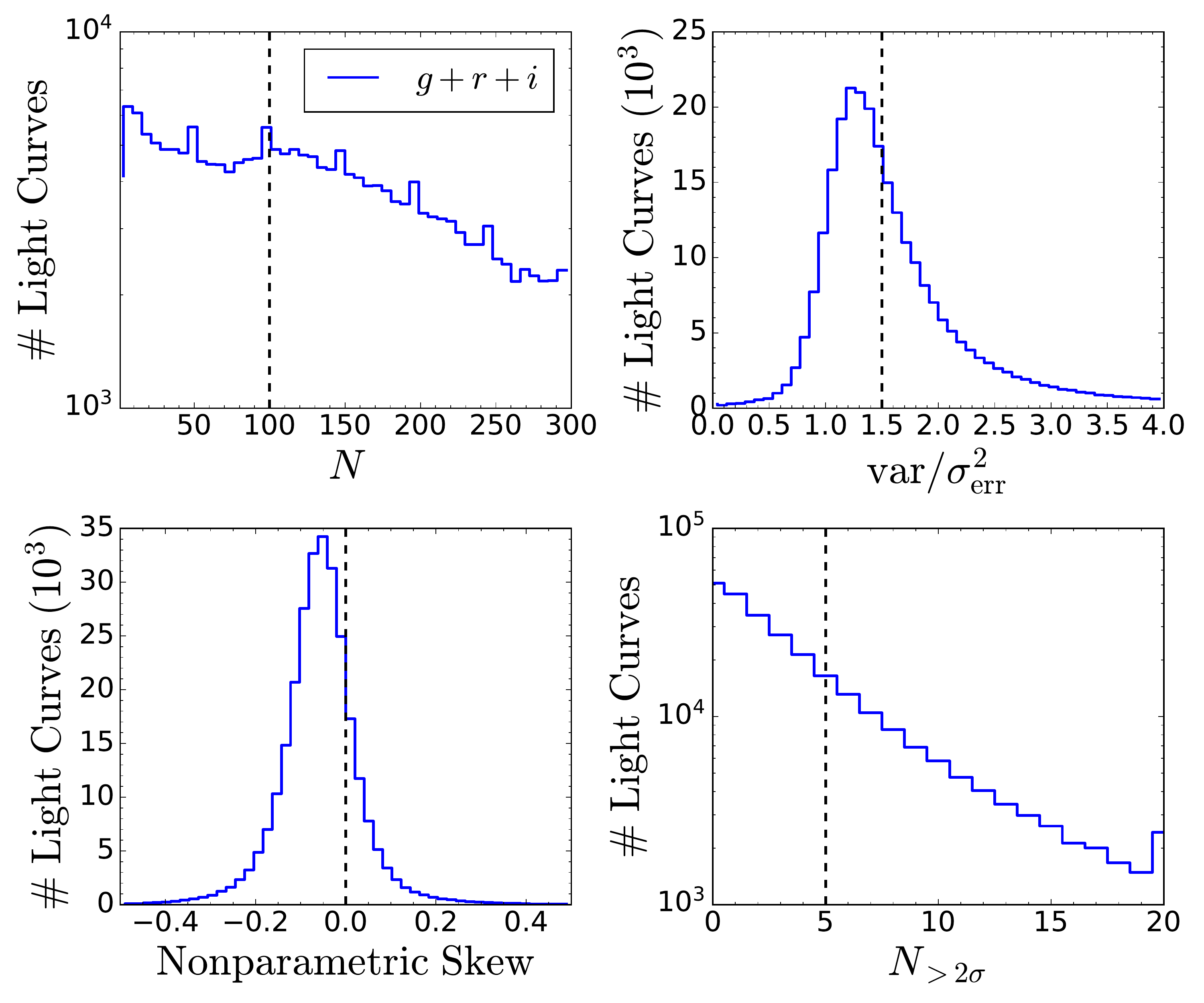}
	\caption{The dashed lines indicate the parameter values above which light curves were selected for the BLS search. The upper left plot shows the histogram of the number of ZTF epochs $N$ contained in a light curve. The upper right plot shows the histogram of the variability metric. The bottom left histogram shows the distribution of nonparametric skew. In the bottom right histogram the statistic $N_{>2\sigma}$ is the number of ZTF epochs with magnitudes that are $2\sigma$ fainter than the median magnitude brightness.
	\label{fig:stat3}}
\end{figure}

The observations in the different filters of ZTF are not simultaneous, so a source typically has three separate light curves, one in each filter. In order to improve the temporal sampling of the light curves when we search for eclipses, we combined $g$, $r$ and $i$ data by matching their median brightness to the $r$-band median magnitude.

Although BLS is a reliable method for detecting eclipses, it is computationally expensive. Running the BLS algorithm on all of the white dwarfs in the catalogue would take a prohibitively long time, so we first reduced the sample using a number of other selection criteria, summarised in Figure \ref{fig:stat3}. 
Firstly, we required that the combined $gri$ light curve contains at least 100 epochs in order to ensure that eclipses are covered by enough data points. This criterion is met by 232\,956 sources. Secondly, the overall variability (i.e. variance) in the light curve should be greater than the mean of the squared magnitude errors in order to eliminate all light curves with no measurable intrinsic variability. The lower limit of the ratio between the variance and the mean of the magnitude errors was set to 1.5. Together with the first criterion this is satisfied by 117\,246 sources. Thirdly, the skewness should be positive, because eclipses are expected to skew the brightness distribution towards fainter magnitude values. As a measure of skewness the nonparametric skew\footnote{The nonparametric skew is the difference between the mean and the median of a light curve normalised by the standard deviation. The mean is more sensitive to outliers than the median. Hence, the magnitudes of a light curve with predominantly positive outliers will have a positive skew.} was chosen, because it is less sensitive to outliers compared to moment skewness. This criterion reduces the set to 27\,268 sources. Lastly, a lower limit was set on the number of ZTF epochs which have magnitudes lying $2\sigma$ above the median magnitude brightness. This lower bound was set to 5 ZTF epochs and ensures that enough data points lie in the eclipses. The subset thus obtained contains 16\,916 sources. The BLS period search was run on this set. In the interest of minimising the computational run-time, we took advantage of multiprocessing, running our search on 32 cores. The computational time for our reduced data set was in the order of 1 week.

In the BLS search, we explored periods ranging from 0.004\,days (5.76\,min) to 0.1\,days (144\,min). They were placed on a uniform frequency grid with a spacing %
\begin{equation}
    \delta f = \frac{\min{(t_\mathrm{E})}}{\max{(P)} \cdot t_\mathrm{B}}
\end{equation}
where $t_\mathrm{E}$ is the eclipse duration, $P$ is the period and $t_\mathrm{B}=\max{(t)}-\min{(t)}$ is the time baseline of the light curve. The period range was chosen taking several factors into account. On the one hand, periods longer than 0.1\,days were avoided because of the accumulation of spurious detections in this period range, likely due to the ZTF sampling pattern. Furthermore, it follows from Kepler's third law that the probability of seeing an eclipse in a given binary system with fixed masses scales as $P^{-2/3}$. The phase span of an eclipse obeys the same scaling relation. %
Hence, a significant detection of a system with fixed binary components becomes less likely at longer periods. %
The computational cost of BLS increases towards shorter periods where the period grid is finer. Therefore, even though ZTF would be able to resolve shorter periods, the lower period limit was set to 5.76\,min, which is somewhat below the period (6.91\,min) of the most compact known eclipsing white dwarf binary.

The BLS search was performed in two separate runs, one run for periods between 0.004\,days and 0.01\,days and another for periods between 0.01\,days and 0.1\,days. This was done in order to allow for different ranges of eclipse durations to be tested, i.e. to test longer durations at longer periods. The eclipse durations $t_E$ tested in the BLS search were placed on an equally spaced logarithmic grid with $\mathrm{d}\log_{10}{t_E}$ = 0.1 ranging from 0.0005\,days to 0.003\,days in the first run and 0.0005\,days and 0.008\,days in the second run. The lower limit, corresponding to 43.2\,s, was set to only allow eclipses that can be resolved within the 30\,s exposure time of ZTF. A sensible upper limit was estimated by calculating the longest possible eclipse duration for two white dwarfs with typical radii orbiting each other at the maximum period, $P=0.01$\,d or $P=0.1$\,d, of the corresponding BLS run. 

The results of the BLS searches were used to further reduce the set of binary candidates. Firstly, light curves were discarded for which a physically impossible eclipse widths, exceeding 0.5 in phase, were found. Secondly, periods of about 0.02\,days and 0.1\,days were identified to accumulate spurious detections. Hence, all light curves were removed for which a period contained in a region of 0.001\,days centred around the aforementioned periods was found. Lastly, a lower limit was set on how many ZTF epochs should be contained in the transit of a BLS result. This limit was set to five and eliminated many detections for which the BLS transit consisted of a single data point. This analysis reduced the results to a set of 1458 binary candidates of which the phase folded light curves were subsequently inspected visually to discern whether they feature eclipses. 

For a small number of these light curves, the strongest period detected in the $P<0.1$\,d range was a harmonic of the true orbital period. The visual inspection revealed that folding the light curve on twice or even four times the value of the strongest period resulted in a much better representation of the eclipse light curve. These are systems with clear secondary eclipses or strong ellipsoidal variation. There are therefore some binaries in our list of detections with periods longer than the nominal period range searched (see Table~\ref{tab:targets}). 

\section{Results}
\label{sec:results}

Using the BLS method described in the previous section, we found nine known eclipsing binaries as well as 18 new binaries that are so far unpublished or have not previously been identified as binary candidates. Their basic \textit{Gaia} properties and orbital periods derived from the ZTF light curves are shown in Table~\ref{tab:targets}. %
For convenience, we also include Table~\ref{tab:phot} in the appendix, which contains their optical and ultraviolet magnitudes. %
The properties of the known binaries recovered are given in Table~\ref{tab:knownbinaries}.

\begin{table*}
\caption{Properties of the eclipsing binaries identified in this search, arranged by increasing orbital period. The $G$ band absolute magnitude was calculated using the distances derived by \citet{bj18}. The typical uncertainty on $T_0$ is $10^{-5}$ days and on $P$ is $10^{-7}$ days. An asterisk in the last column indicates an uncertain classification, because no clear features from the companion star were detected in the spectrum. \label{tab:targets}}
\renewcommand{\arraystretch}{1.3} 
\begin{tabular*}{\textwidth}{l| @{\extracolsep{\fill}} cccccccccccccccccc}
\hline\hline
\multicolumn{1}{c|}{ZTF Name} & Gaia DR2 & RA & DEC & $\omega$ & $G_\mathrm{BP}$--$G_\mathrm{RP}$ & $G_\mathrm{abs}$ & $P$ & $T_0$ & Class \\
~ & sourceID & (deg) & (deg) & (mas) & (mag) & (mag) & (days) & (BMJD(TDB)) & ~\\\hline
  J1637+4917 & 1410739870171621504 & 249.43121 & 49.294556 & 4.58$\pm$0.27 &  0.27  &  12.65$^{+0.13}_{-0.14}$ &  0.04270769 &  58198.46416 &  AM\,CVn \\
  J0219+4623 & 353846172081048960  & 34.832417 & 46.395863 & 1.84$\pm$0.54 &  0.47  &  10.87$^{+0.78}_{-1.16}$ &  0.05832248 &  58283.45225 &  CV \\
  J2212+5347 & 2004624931143291648 & 333.11234 & 53.797417 & 5.14$\pm$0.17 &  0.41  &  11.85$^{+0.07}_{-0.07}$ &  0.05833898 &  58217.52413 &  WDMS$^*$ \\
  J1356+5705 & 1657868410028294784 & 209.11125 & 57.095917 & 2.69$\pm$0.22 &  0.24  &  11.10$^{+0.18}_{-0.19}$ &  0.06383960 &  58198.40872 &  ELM \\
  J1747+5529 & 1418364678367782016 & 266.94693 & 55.488277 & 2.44$\pm$0.26 &  0.58  &  11.09$^{+0.23}_{-0.26}$ &  0.06388127 &  58198.4429  &  CV \\
  J2048+1930 & 1814387802956119552 & 312.18622 & 19.505388 & 0.72$\pm$0.39 &  0.16  &  8.57$^{+1.07}_{-1.40}$  &  0.06486074 &  58211.55589 &  CV \\
  J1758+7642 & 1656197800893107712 & 269.55353 & 76.704666 & 1.81$\pm$0.23 &  -0.18 &  10.25$^{+0.28}_{-0.32}$ &  0.06566667 &  58198.41165 &  DWD/ELM \\
  J0609+3652 & 3453995450857025664 & 92.309875 & 36.867474 & 1.11$\pm$0.34 &  -0.06 &  8.87$^{+0.80}_{-1.18}$  &  0.06604491 &  58204.14731 &  WDMS$^*$ \\
  J1100+5210 & 836510989033113472  & 165.18808 & 52.17878  & 1.90$\pm$0.22 &  -0.21 &  9.94$^{+0.25}_{-0.28}$  &  0.06688196 &  58202.33483 &  WDBD$^*$ \\
  J1513+7037 & 1694471835016084608 & 228.38737 & 70.62289  & 2.06$\pm$0.32 &  0.5   &  10.99$^{+0.34}_{-0.40}$ &  0.06882375 &  58198.37066 &  CV \\
  J1844+4857 & 2119978952315202176 & 281.14328 & 48.96014  & 1.31$\pm$0.29 &  0.18  &  9.83$^{+0.52}_{-0.67}$  &  0.07266928 &  58198.44812 &  WDMS$^*$ \\
  J1644+2434 & 1300187622427241856 & 251.17163 & 24.5745   & 2.40$\pm$0.29 &  0.19  &  10.97$^{+0.27}_{-0.30}$ &  0.08010533 &  58198.44721 &  WDMS$^*$ \\
  J2036+5506 & 2183626416274746496 & 309.00696 & 55.11353  & 1.12$\pm$0.54 &  -0.08 &  8.84$^{+1.59}_{-2.39}$  &  0.08155910 &  58205.51196 &  sdMS \\
  J2219+5628 & 2006245366462452096 & 334.80304 & 56.468166 & 0.84$\pm$0.43 &  0.29  &  8.71$^{+1.27}_{-1.79}$  &  0.08795522 &  58217.57416 &  WDMS$^*$ \\
  J1724+6044 & 1438919253678919040 & 261.00015 & 60.747723 & 1.05$\pm$0.57 &  0.15  &  10.10$^{+1.00}_{-1.27}$ &  0.08872902 &  58198.46341 &  WDMS$^*$ \\
  J2131+4654 & 1978219536621659136 & 322.92267 & 46.908333 & 0.56$\pm$0.10 &  0.03  &  6.19$^{+0.38}_{-0.45}$  &  0.12996811 &  58208.55693 &  sdMS \\
  J1920+2722 & 2025873096433233664 & 290.05887 & 27.371695 & 5.22$\pm$0.04 &  0.37  &  9.16$^{+0.02}_{-0.02}$  &  0.14936800 &  58216.63748 &  WDMS \\
  J1936+3710 & 2048949573450878336 & 294.02032 & 37.171444 & 0.40$\pm$0.13 &  -0.17 &  5.91$^{+0.73}_{-1.03}$  &  0.23834994 &  58204.72766 &  sdMS 

\\\hline
\end{tabular*}
\end{table*}

\begin{figure}
    \includegraphics[width=\linewidth]{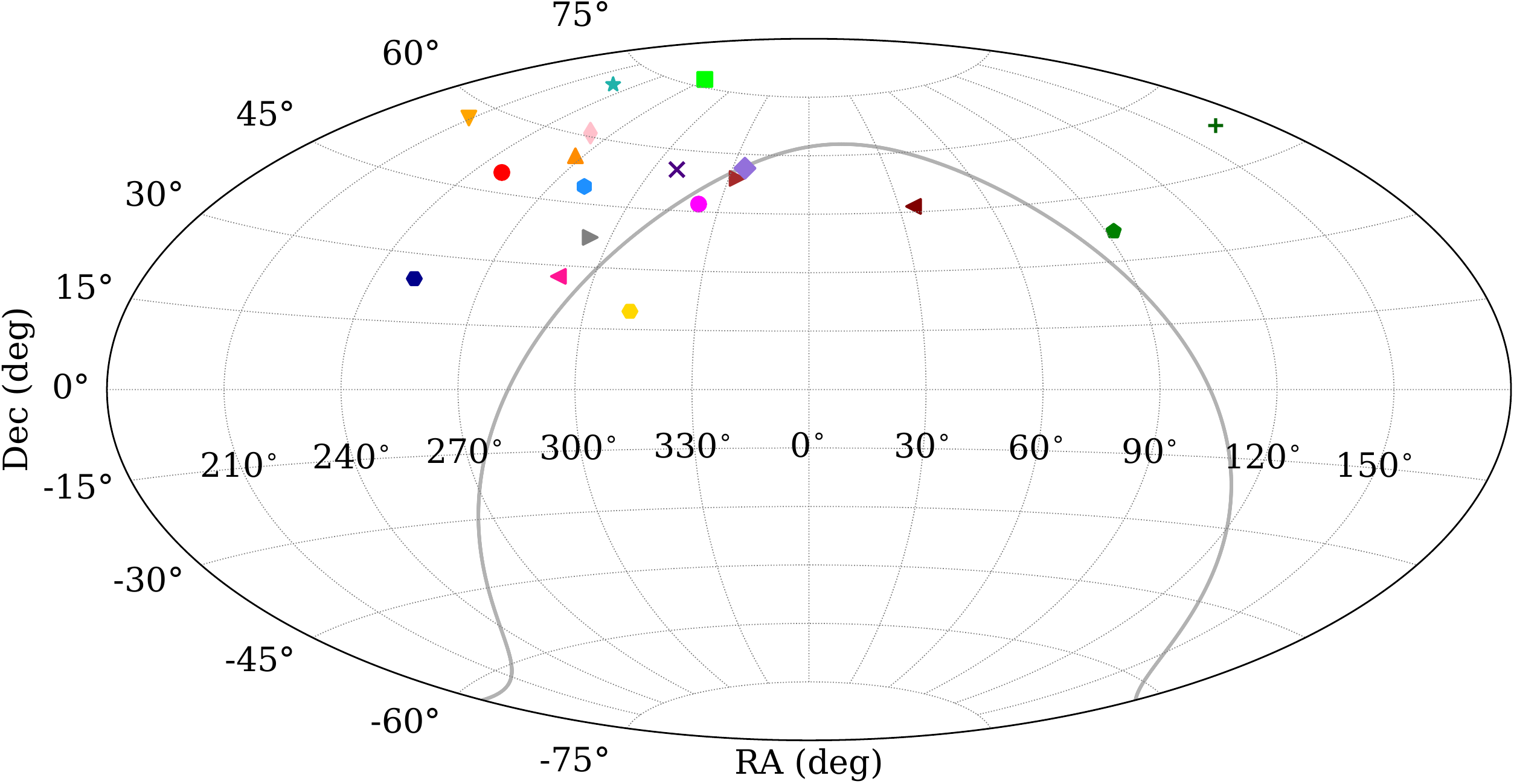}
	\caption{The newly detected eclipsing binaries, plotted in equatorial coordinates. The symbols are as defined in  Figure~\ref{fig:HR_eclipses}. The solid grey line represents the Galactic plane.
	\label{fig:skyplot}}
\end{figure}

We start with a quick summary of positional and kinematic properties of our sample. Figure~\ref{fig:skyplot} gives the distribution on the sky of the newly discovered binaries  in equatorial coordinates. As expected, the stars congregate in the Northern hemisphere due to the limitations of the ZTF survey footprint. Additionally, most of the binaries seem to prefer the range of $220^{\circ}<\mathrm{RA}<330^{\circ}$. We link this apparent clustering to the details of the ZTF's cadence: the total number of measurements plays an important role in the detection of short-duration eclipses, as elucidated in Section~\ref{sec:sims}. 

\begin{figure}
    \includegraphics[width=\linewidth]{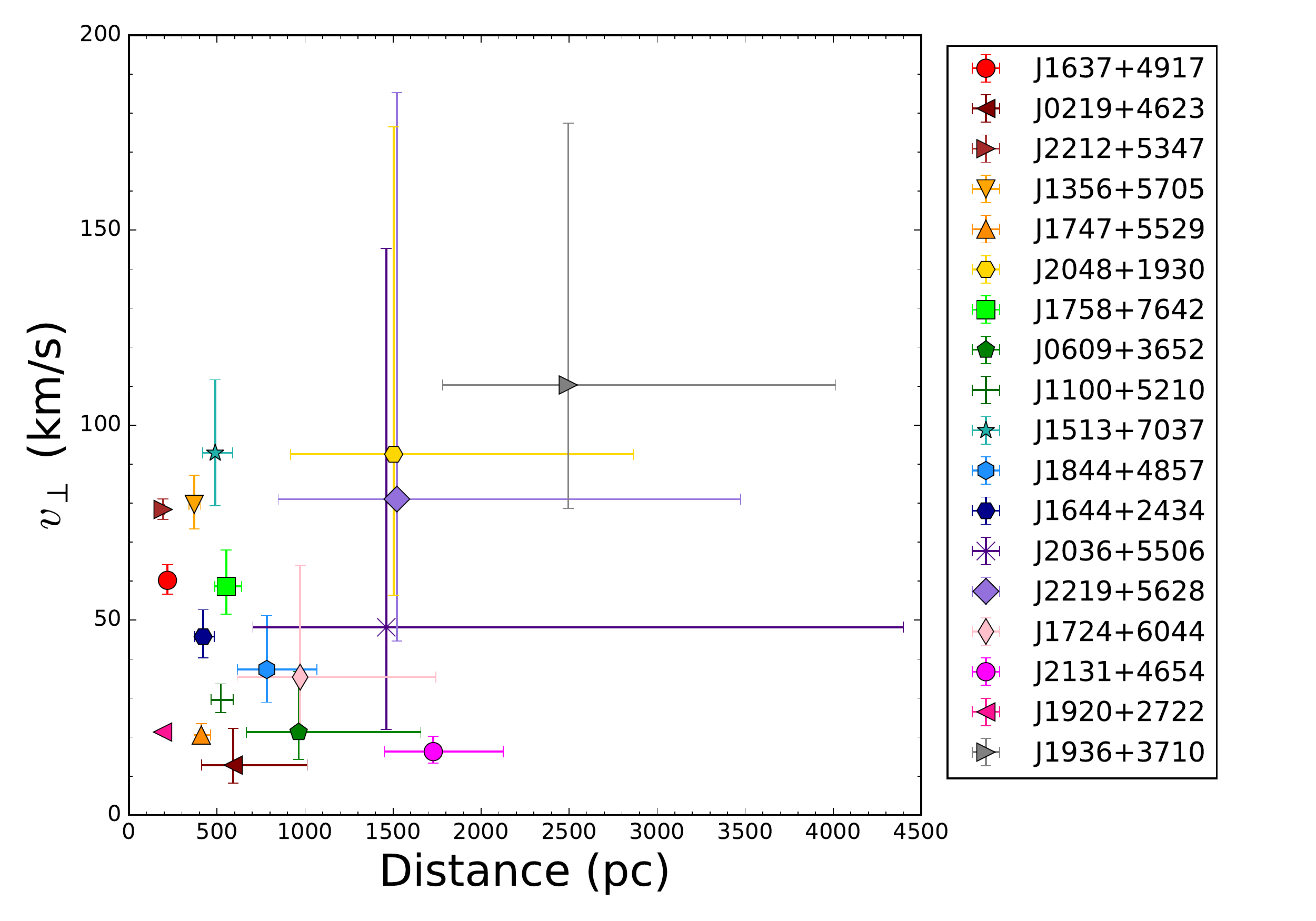}
	\caption{Amplitude of the heliocentric tangential velocity in km\,s$^{-1}$ as a function of heliocentric distance in pc. Elevated tangential velocities together with reasonably large distances point to the thick disc origin for the bulk of our binaries.\label{fig:vt}}
\end{figure}

Figure~\ref{fig:vt} shows the distribution of the heliocentric tangential velocities (in km/s), obtained from the \textit{Gaia} EDR3 measurements of the proper motion amplitude $V_{\rm t}=4.74047D\sqrt{\mu^2_{\alpha}+\mu^2_{\delta}}$, as a function of the heliocentric distance (in pc). %
$D$ is the distance to the source and $\mu_{\alpha},\mu_{\delta}$ the proper motion in RA and Dec, respectively. %
Because most of our unresolved binaries are significantly brighter than single WDs, they cover a substantially broader range of heliocentric distances. \textit{Gaia}'s white dwarf detections are typically limited to 100--300\,pc, while a half of our sample is at distances beyond $\gtrsim550$\,pc. Combined together with a characteristic Galactic latitude of $\sim30^{\circ}$, large heliocentric distances imply significant heights above the plane of the Milky Way's disk. In other words, approximately half of our sample is above $|z|>250$pc, i.e. just over the scale-height of the Galactic thin disc \citep[e.g.][]{Juric2008,Kordo2011}. That the discovered white dwarf binaries are predominantly members of a kinematically ``hot'' and therefore old Galactic population is supported by their tangential velocities (also shown in in Figure~\ref{fig:vt}.) With the tangential velocity dispersion of $\sim26$ km\,s$^{-1}$, the representative age for the binaries in our sample is $\sim8-10$~Gyr, using the age-velocity relation for the $V_z$ component as presented in \citet{Sharma2020}. Given the appreciable Galactic heights and substantial tangential velocities, it is also possible that some (most) of our binaries are denizens of the Galactic thick disc instead.  

\begin{figure*}
    \includegraphics[width=\linewidth]{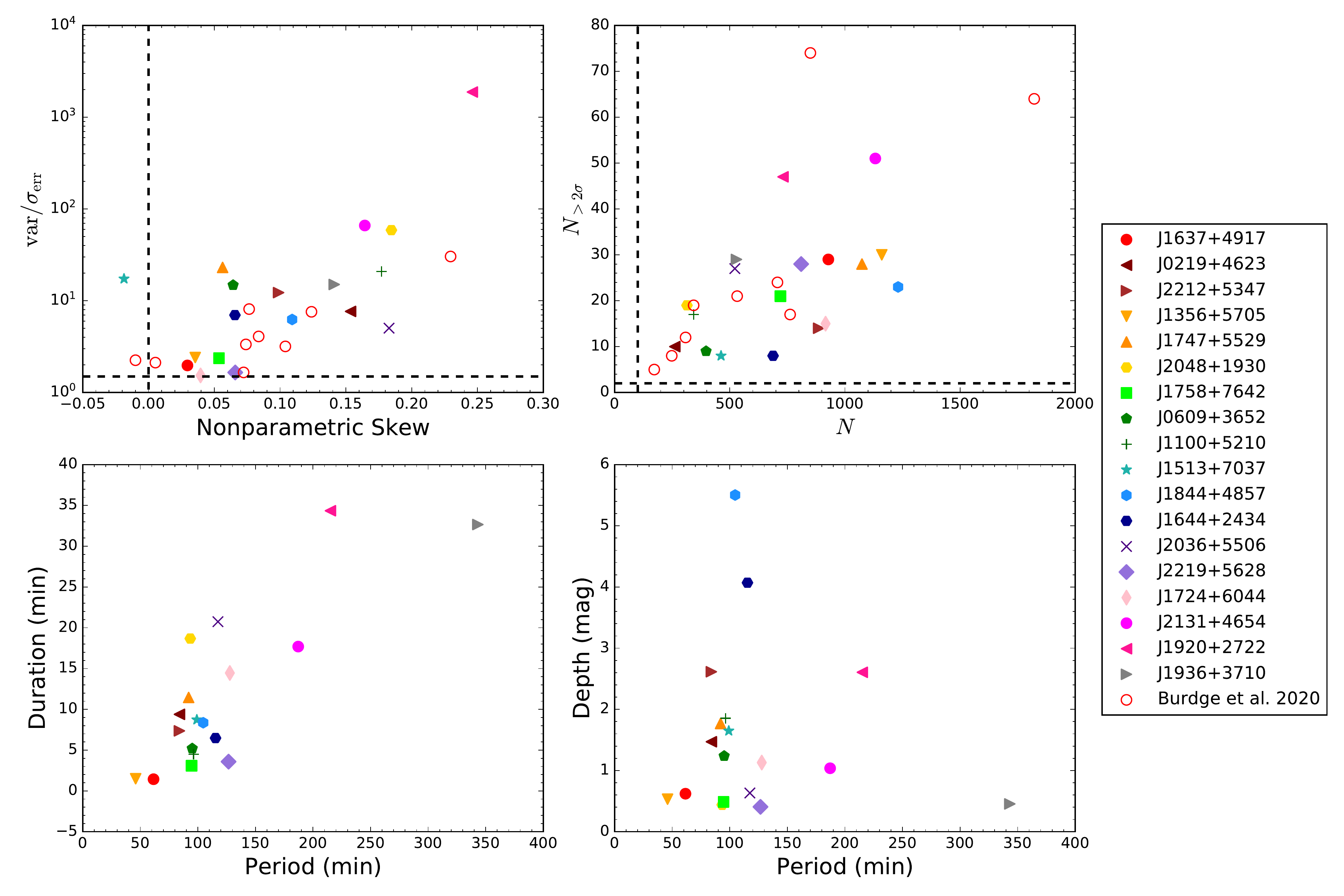}
	\caption{Summary of light curve properties of the new eclipsing binaries. The top two plots show the location of the binaries within the parameter space used for the selection of ZTF light curves prior to the period search (cf. Figure \ref{fig:stat3}). For reference, we have included eclipsing binaries listed in \citet{Burdge20}. Note that in the top left plot there are two sources with negative skew, i.e. below our skewness cut. They are J1513+7037 from our own sample and J1905+3134 from \citet{Burdge20}. Both these sources are disc-accreting white dwarfs with bright outbursts. J1513+7037 is included here, because it was detected in an earlier ZTF data release where it did have positive skew. The bottom plots show the periods, durations and depths of the eclipses obtained by fitting a triweight kernel. \label{fig:summary}}
\end{figure*}

Figure \ref{fig:summary} summarises the light curve properties of the detected binary objects. %
\citet{Burdge20} presented a further nine eclipsing binaries found in the ZTF data, so we include those binaries in the plots as well, to compare their properties with those found by us.
The top two plots show the values of the light curve statistics which were used for the selection prior to the period search (cf. Section~\ref{sec:BLS}). As expected, the eclipsing sources that have a high light curve variability also tend to have a higher positive skew and ones that have more data points tend to have more outliers in the positive magnitude direction which increases the detectability. There are only two sources that have negative skew. One is ZTF\,1513+7037 from our own sample and the other is ZTF\,J1905+3134 from \citet{Burdge20}. These are sources that have strong non-periodic variability producing bright outliers that skew the light curve in the opposite direction of the eclipse. The reason for the inclusion of the negatively skewed source ZTF\,J1513+5210 is that it was originally detected in an earlier data release (ZTF DR2) where it had positive skew. A bright outburst detected during the DR3 data collection period changed the skew from positive to negative. 
Six of the nine eclipsing sources detected by \citet{Burdge20} do not appear in the white dwarf catalogue of \citet{gentile2019}, so they would have been excluded from our search. The remaining three lie above the cuts and were recovered by our search. We discuss these systems further in Section~\ref{sec:completeness}.
The bottom plots of Figure~\ref{fig:summary} show the different periods, durations and depths of the eclipses of our binary candidates, discussed in detail in Section~\ref{sec:targetsdetail}. As expected, objects with longer periods also tend to have longer eclipse durations.

\begin{figure*}
    \includegraphics[width=\linewidth]{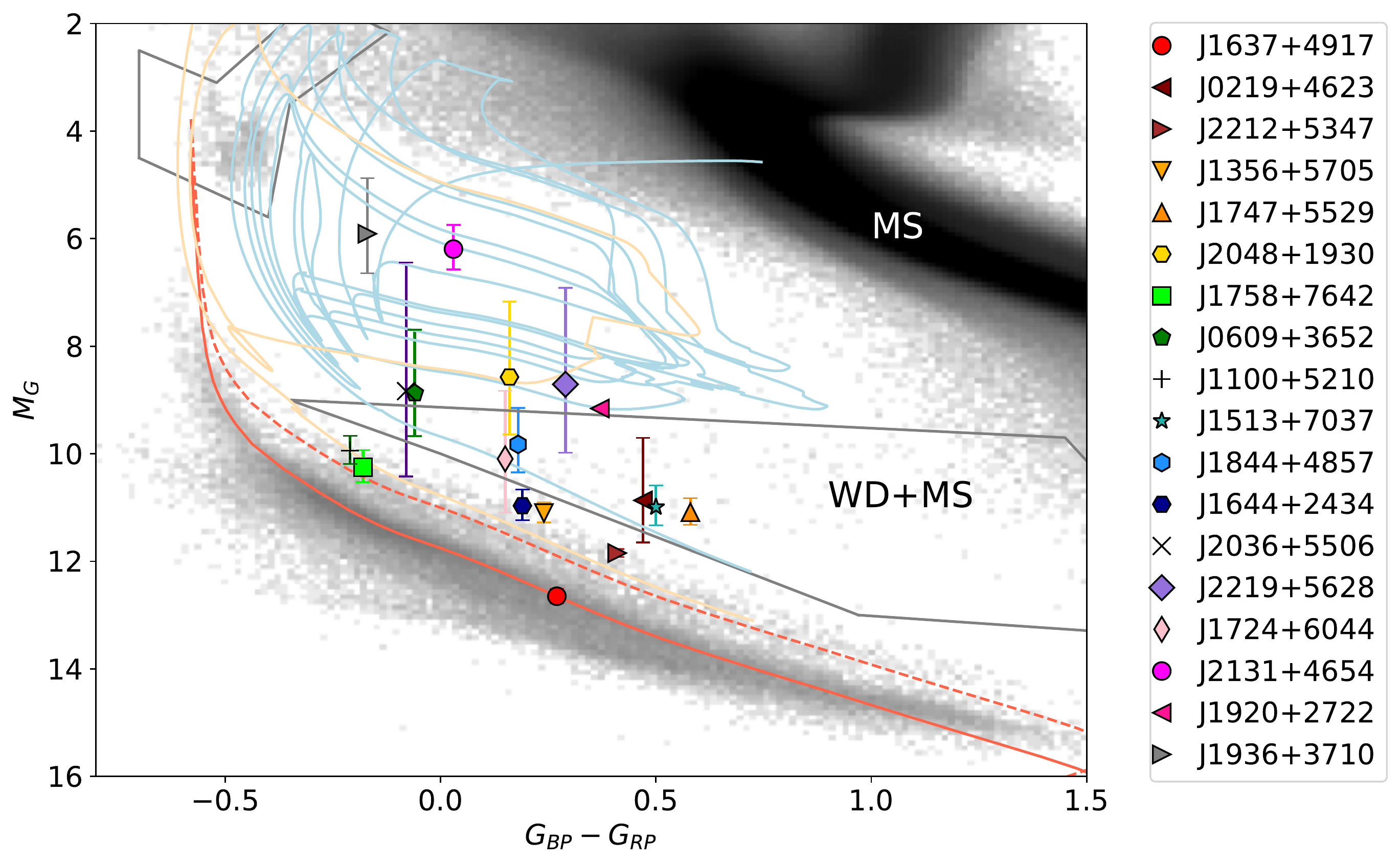}
	\caption{The location of the new eclipsing binaries on the \textit{Gaia} H-R diagram. The background grey-scale 2D histogram shows the logarithm of the density of the nearby sources measured by {\it Gaia} EDR3, i.e. those with $\varpi/\sigma_{\varpi}>20$ and $|b|>20^{\circ}$ (the magnitudes and colours are corrected for dust extinction). Note that for the background sources, distances are simply $\varpi^{-1}$ (a reasonable assumption given their high-quality parallax measurements), but for the eclipsing binary systems, absolute magnitudes were calculated using the distances derived by \citet{bj18} for the \textit{Gaia} DR2 catalogue. Several familiar regions are apparent in the grey-scale density distribution: the sequence occupied by the white dwarfs, the main sequence (MS), the main sequence turn-off and the Extreme Horizontal Branch (top left corner of the H-R diagram, delimited by a black solid line). Also marked is the ``bridge'' composed of white dwarf--main sequence binaries as delineated by \citet{Belokurov2020}.
	In the legend, binaries are arranged by increasing orbital period, as detected by BLS. The solid red line is the cooling track for a $0.6 M_{\odot}$ white dwarf from \citet{Bergeron19}, while the red dashed line is the same model offset by $-0.75$ mag. Note that only a small fraction of our objects (perhaps as many as 6 systems) are located on or close to the main white dwarf sequence, while the rest are significantly brighter occupying the region in and around the white dwarf--main sequence ``bridge''. Finally, some of the eclipsing binaries discovered may contain an ELM white dwarf as indicated by the ELM models show. Light blue (light yellow) track is the ELM model from \citet{Pelisoli19} for the final mass of $0.182 M_{\odot}$ ($0.324 M_{\odot}$).
	\label{fig:HR_eclipses}}
\end{figure*}

The eclipsing binaries we detected could be of various types, including double white dwarfs, white dwarf and main sequence star binaries, cataclysmic variables and hot subdwarf binaries. 
A full classification with low cadence photometric data, such as ZTF, is not always possible. However, with the addition of \textit{Gaia} and other auxiliary data it is possible to make some constraints or to rule out certain classes of sources \citep[e.g.][]{Eyer19,Inight21}. 
The 18 new binaries are shown on the \textit{Gaia} H-R diagram in Figure~\ref{fig:HR_eclipses}, along with various evolutionary tracks and regions of interest. Firstly, we show a representative white dwarf cooling track as a solid red line. This model corresponds to a $0.6 M_{\odot}$ white dwarf \citep{Bergeron19}\footnote{White dwarf evolutionary sequences for various masses are available from \url{https://www.astro.umontreal.ca/~bergeron/CoolingModels/}}.
The red dashed line is the same model offset by 0.75\,mag, to indicate the location of double white dwarfs consisting of two $0.6 M_{\odot}$ stars. The ELM white dwarfs have very different paths across the H-R~diagram, depending on the mass of the hydrogen envelope and the number of hydrogen shell flashes it undergoes \citep{Istrate16}. Figure~\ref{fig:HR_eclipses} shows the tracks of ELM white dwarfs with final masses of $0.182 M_{\odot}$ and $0.324M_{\odot}$ in light blue and yellow, respectively \citep{Pelisoli19}. Finally, we also indicate the region at the top left of the diagram where the Extreme Horizontal Branch stars are found, as well as the ``bridge'' of white dwarf--main sequence binaries, described by \citet{Belokurov2020}. Note that white dwarf--main sequence binaries are found outside this region as well. This region describes the highest concentration of these stars in the H-R diagram, derived from the \textit{Gaia} astrometric variability parameters (and validated using known systems). 

Overall, the majority of the eclipsing white dwarf binaries detected here lie well above the white dwarf sequence. Their location in the H-R diagram towards the ``bridge'' suggest that main sequence companions are the most likely, but ELMs and subdwarfs are possible too. In Section~\ref{sec:targetsdetail} we consider each system individually, along with the information derived from their ZTF light curves. 

\section{Discussion}
\label{sec:discussion}

\subsection{Completeness}
\label{sec:completeness}

This search for eclipsing white dwarf binaries revealed a variety of binary types across a range of periods. In addition to the 18 new or unpublished systems shown in Figure~\ref{fig:HR_eclipses}, our method also detected nine known or recently-identified systems. They are the double white dwarf binaries ZTF\,J190125.42+530929.5 \citep{coughlin2020}, ZTF\,J053802.71+195303.0 and ZTF\,J072221.49-183930.6 \citep{Burdge20}, the fully-eclipsing AM\,CVn star Gaia14aae \citep{Campbell15,Green18}, the short-period cataclysmic variables WZ\,Sge \citep{Patterson18}, OV\,Boo \citep{Uthas11}, NZ\,Boo \citep{Littlefair08} and SDSS\,J093537.46+161950.8 \citep{Southworth15}, as well as the sdB--M-dwarf binary PTF1\,J011339.09+225739.1 \citep{Wolz18,chen20}. For completeness, the properties of these binaries are shown in Table~\ref{tab:knownbinaries} of the appendix.

However, it is also notable that there are some known systems which were not detected in this initial search, because of the target selection method we chose. Our method requires the target to be included in the \textit{Gaia} white dwarf catalogue, but that automatically excludes binaries which did not have high enough quality parallaxes in \textit{Gaia} Data Release 2 (as defined by \citealt{gentile2019}). 
For instance, six of the nine eclipsing binaries listed in \citet{Burdge20} are not contained in the \textit{Gaia} white dwarf catalogue (see Table~\ref{tab:notinwdcat} for details), so would not be recovered by our method.
In addition, one of those six binaries, ZTF\,J1905+3134, would also have been rejected because of its negative skew in the ZTF DR3 data, so it is possible that there are other systems which were similarly lost to the initial cuts we applied in Section~\ref{sec:BLS}.

It is difficult to make a reliable estimate of the completeness of our search, since no complete catalogue exist to compare against. For a rough estimate, we can compare the binaries we found to the known eclipsing binaries in the International Variable Star Index database, VSX\footnote{VSX is operated by the American Association of Variable Star Observers, Cambridge, Massachusetts, USA, and is available online at \url{http://www.aavso.org/vsx/}}. 
We conducted a thorough literature search to compile a list of known eclipsing white dwarf binaries in the period range we searched, but all of the binaries we found were already included in VSX. So, we believe that VSX provides a good record of the known systems. First, we have to determine how many eclipsing white dwarfs are known in the area covered by ZTF. Since the targets in VSX are listed at J2000 coordinates, we first rolled back the \textit{Gaia} J2015.5 coordinates of the white dwarf catalogue to J2000, then cross-matched the white dwarf catalogue to VSX using a 2$\arcsec$ radius. To match the ZTF public survey area, we selected all stars north of Dec$>-31^{\circ}$ \citep{bellm2018}, and then select all that are flagged as confirmed eclipsers. Finally, we select only those with orbital periods matching our search range, i.e. $P\leq144$\,min. This gives 49 known eclipsers in this sky area, of which we recovered nine. 

Of the 40 known eclipsing binaries not recovered, we found that all could be explained within our search criteria and the available ZTF data. Nine of the 40 have no ZTF data and a further eight have poor sampling (fewer than 10 epochs) so were rejected due to our requirement of having 100 points in the light curve. Another eight have fewer than five points sampling the eclipse, and for three binaries the eclipse is not detectable in the ZTF light curve. 10 of the binaries are cataclysmic variables with large amplitude outbursts that skewed their light curves to negative values. The remaining two have deep, relatively well-sampled eclipses in the most recent ZTF data release, but at the time of our search, using DR3, there were not enough points in their eclipses to be included.

These results show that the internal completeness of our search is high, i.e. the non-detection of known systems can be fully explained within the search criteria we applied. However, it is very likely that many more binaries remain to be found in the ZTF data release,
and that the external completeness is much lower. Without a statistically complete catalogue to compare against, it is not possible to assign a numerical value to the overall external completeness. A first step to recover more of the known binaries, and potentially new binaries with similar properties, would be to use different or more relaxed search criteria than we applied here. Note for example, that since we used the white dwarf catalogue in the cross-match, this completeness estimate also excludes any binaries missed because they were not included in that catalogue. 

Three of the newly discovered eclipsing binaries (ZTF\,J2131+4654, ZTF\,J1936+3710 and ZTF\,J1920+2722) were initially detected at a harmonic, such that the true orbital period was longer than our search limit of 144\,min. This highlights that exploring periods longer than 0.1~days will yield more detections of interesting, although less compact, binary systems. 
With the increasing amount of ZTF data, further short period systems and double degenerate binaries are also expected \citep[e.g.][]{burdge2019,Burdge20_9min}. In our method, more detections could be achieved by lowering detection thresholds and exploring larger subsets of the data. Our search is ongoing, and with an increasing number of epochs in the public data releases, we expect to find more of these systems.

\subsection{Detection Efficiency Simulations}
\label{sec:sims}

\begin{figure*}
	\centering
		\includegraphics[width=\textwidth]{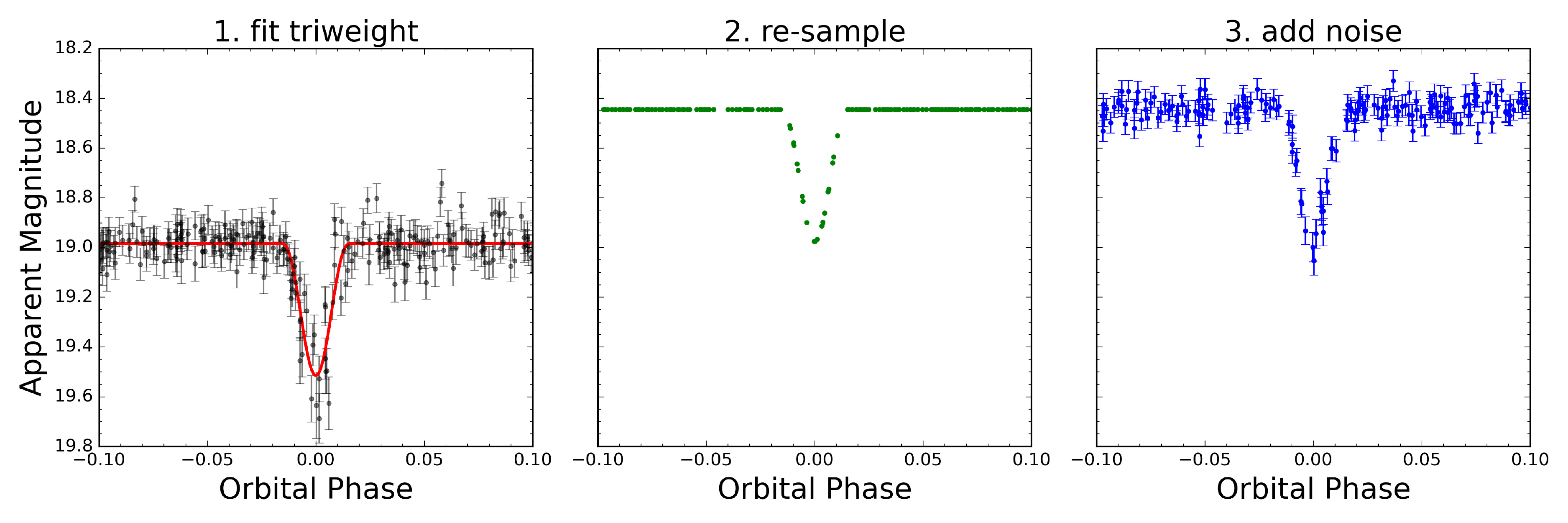}\\
	\caption{Each eclipse is fitted with a triweight kernel (step 1) which is used to generate a mock light curve with the sampling and mean magnitude of a random ZTF source (step 2). Lastly noise is added that depends exponentially on magnitude (step 3). The period search is then run on this mock light curve to determine whether it would be detected with the BLS method and selection criteria used in this work. This is repeated 1000 times for each source to calculate the detection efficiency.
	\label{fig:mocklc}}
\end{figure*}

Given the complexity of a full completeness analysis of our survey, we estimate here the detection efficiency on a per-object basis. We find that the number of epochs in the light curve as well as the phase width of the eclipse are the crucial parameters in detecting the eclipse and measuring the orbital period of the binary accurately. 

For each of the eclipsing sources detected at their real period (i.e. non-harmonics with $P<0.1$\,d), we ran simulations in order to determine the efficiency with which similar binaries were detected in the ZTF data. In particular, we wanted to test the effect of variable sampling on the detection of the eclipse. We generated mock light curves by fitting a model to each of the eclipse light curves. The box model derived from BLS is a poor representation of a real eclipse and hence alternative kernels were tested to model the eclipse shape. Overall, the triweight kernel %
\begin{equation}
    K(\phi) = d \left(1 - \left(\frac{2 \phi}{w}\right) ^2\right)^3
\end{equation}
was found to yield the best fits of the various eclipses. Here $d$ and $w$ are the eclipse depth (in magnitude) and phase width respectively and $\phi$ is the orbital phase. Using the best-fit kernel parameters, 1000 mock light curves were generated for each source using the real sampling and mean magnitudes of a randomly selected sample of ZTF sources containing more than 50 epochs. Noise was added to the epochs using a noise model that depends exponentially on magnitude. The noise model parameters were determined by fitting the model to randomly selected ZTF epochs covering a wide range in brightness. The procedure is illustrated in Figure~\ref{fig:mocklc}.

The BLS method was run on the mock light curves with the same settings as described in Section \ref{sec:BLS}. A mock eclipse light curve was deemed detected when the BLS period equaled the real period within a relative tolerance of $10^{-3}$. Moreover, the mock light curve was required to pass the same selection criteria as the real light curves (Figure \ref{fig:stat3}). The only selection criterion which was relaxed is that of variability. This choice was made, because most detected light curves featuring eclipses have additional variability (e.g. outliers and ellipsoidal variations) which is not incorporated in the model and would thus lead to an underestimate of the detection efficiency. For each source in Section~\ref{sec:results} we report the total efficiency $E_\mathrm{tot}$ which is defined as the fraction of all  mock light curves with more than 50 epochs (i.e. $N>50$)
in which the eclipse was detected.

\subsection{Notes on individual objects}
\label{sec:targetsdetail}

\begin{figure*}
	\centering
	\includegraphics[width=0.85\linewidth]{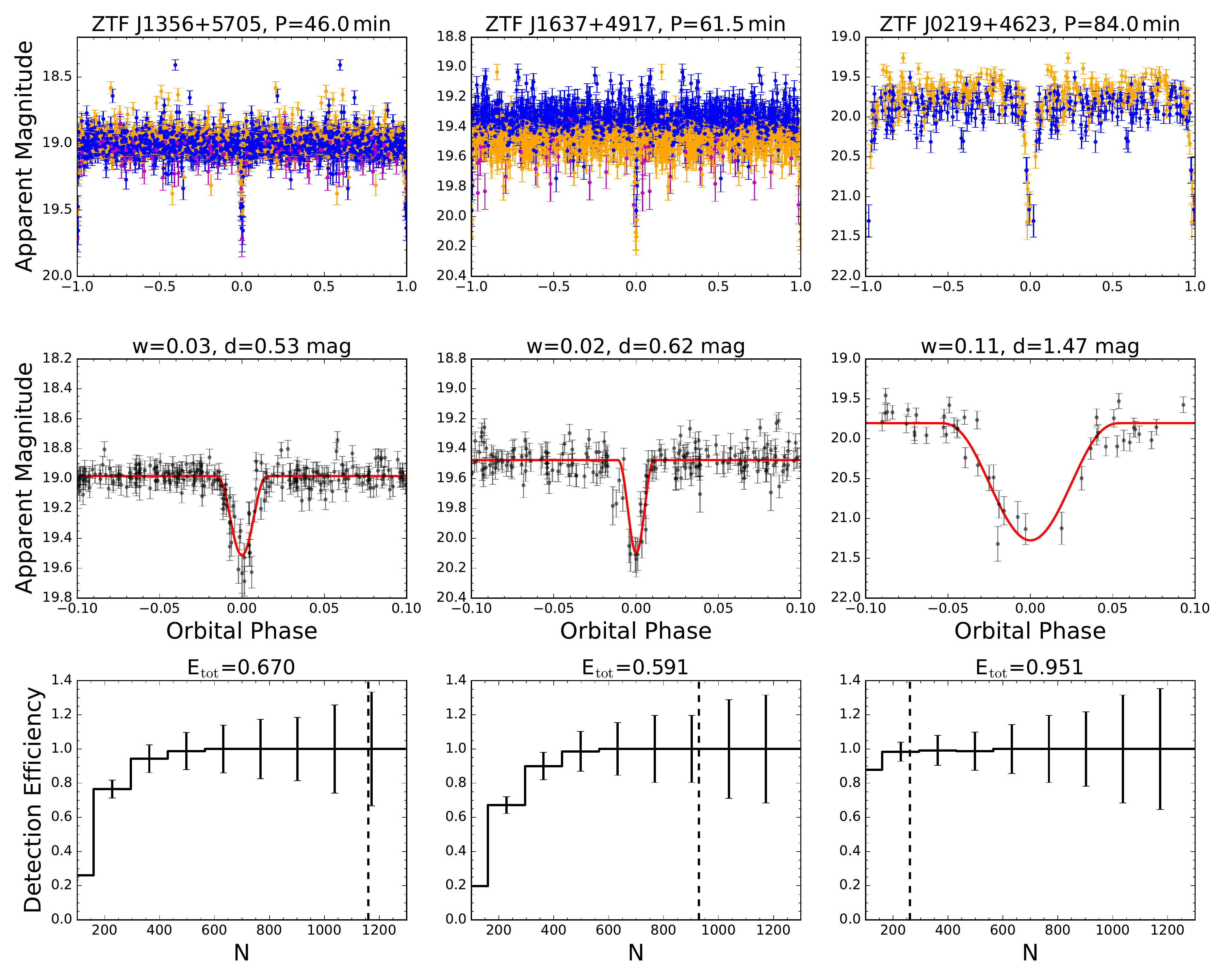}
	\caption{The three figure columns correspond to the three shortest period binaries detected by our BLS search. The top row shows the phase folded light curves where the orange markers denote epoch measured in the $r$-band, the blue markers denote epochs measured in the $g$-band and the magenta markers, if the data is available, denote the epochs measured in the $i$-band. The middle row shows the triweight fits of the eclipses. The best-fit width ($w$) and depth ($d$) of the eclipse are given in the plot titles. The bottom row shows the results of the detection efficiency simulations, as a function of the number of epochs $N$ in the light curve. $E_\mathrm{tot}$ denotes the total efficiency of detecting the source for $N>50$ and the dashed line indicates the number of data points contained in the original light curve. ZTF\,J1356+5705 and ZTF\,J1637+4917 have similar light curve features. Due to their narrow and shallow eclipses they are poorly detected at small $N$ in the simulation. ZTF\,J0219+4623, on the other hand, has a much broader and deeper eclipse and is thus well detected across all $N$ in the simulation. Follow-up observations have shown that the orbital period of ZTFJ1356+5705 is double that reported here (see text).}
	\label{fig:eclipse1}
\end{figure*}

\paragraph*{ZTF J135626.70+570545.3} (J1356+5705) has the shortest period as revealed by our BLS search. The ZTF light curve, folded on a period of 45.96\,min, is shown in Figure~\ref{fig:eclipse1}, along with the triweight kernel fit to the eclipse. 

\begin{figure*}
	\centering
		\includegraphics[width=0.85\linewidth]{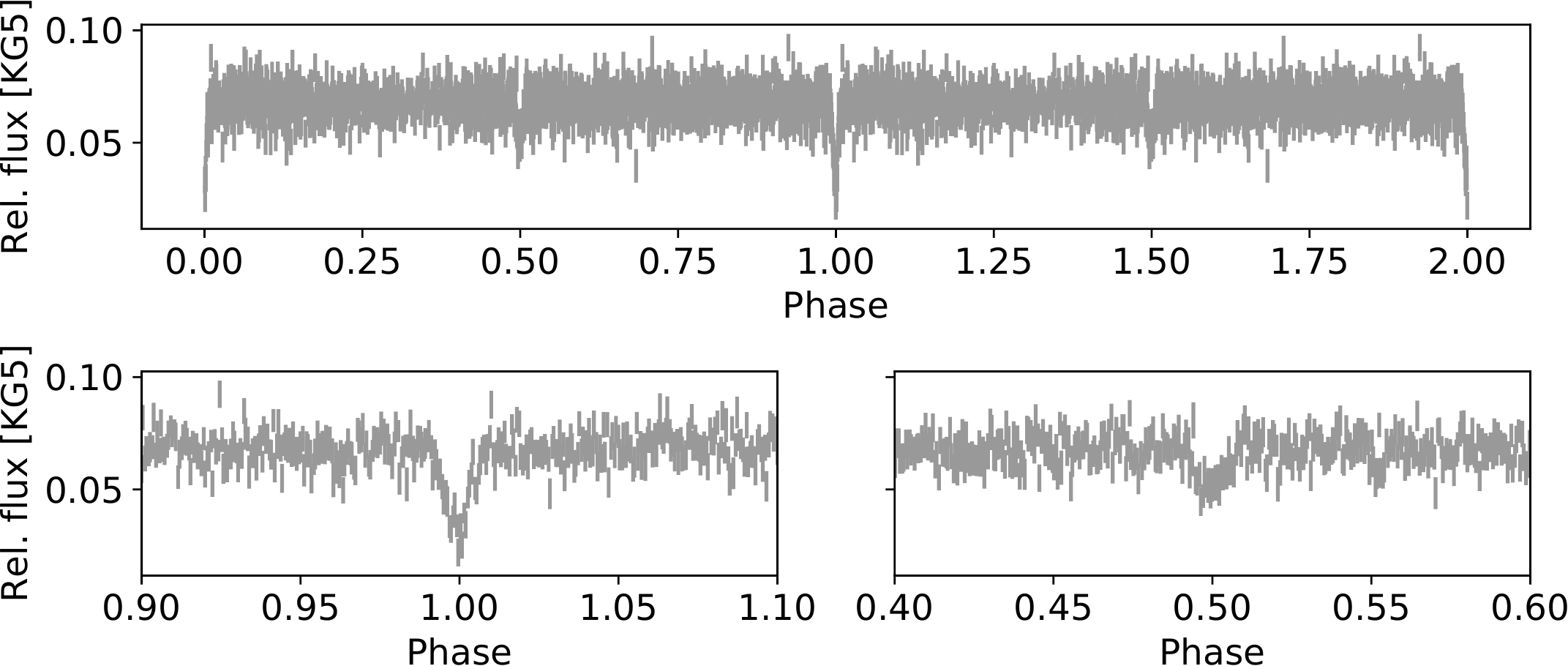}
	\caption{ULTRASPEC light curve of ZTF\,J1356+5705, folded on the ephemeris in Table~\ref{tab:targets}. The bottom panels show the regions around the primary and secondary eclipses.
	\label{fig:J1356uspec}}
\end{figure*}

We observed J1356+5705 on 2021 April 9 using the high-speed photometer ULTRASPEC \citep{Dhillon14} on the 2.4\,m Thai National Telescope at Doi Inthanon, Thailand. Conditions were moonless but cloudy, and we measured the median seeing to be 1.7\,$\arcsec$. We obtained a total of 1535 6.35\,s exposures over 2.7\,hours in the $KG5$ filter. This non-standard blue filter is useful for faint targets like these, because of its wide throughput. The filter is described in detail by \citet{Hardy17}. The light curve, shown in Figure~\ref{fig:J1356uspec}, revealed distinctly different depths for alternate eclipses, implying a primary and secondary eclipse. The orbital period of the binary is therefore twice that detected with BLS, $P=91.93$\,min. 

With $G_\mathrm{BP} - G_\mathrm{RP} = 0.24$\,mag, J1356+5705 is relatively red in colour and it lies about 1\,mag above the white dwarf sequence at $G_\mathrm{abs} = 11.10$. The probability of being a white dwarf \citep{gentile2019} is $P_\mathrm{WD}=0.93$ which suggests that there is a detectable contribution from the binary companion. %
The spectrum of this object (Figure~\ref{fig:spectra}) shows deep, relatively narrow, Balmer absorption lines on a blue continuum, and identifies it as a likely extremely low mass (ELM) white dwarf binary \citep{Pelisoli19}. This is also consistent with its position on the H-R diagram just above the white dwarf sequence and among the ELM tracks. Further spectroscopic and high-speed photometric observations will be carried out to measure the binary parameters and confirm this classification. In the simulation J1356+5705 is nearly always detected in light curves with $N>400$. In light curves with fewer epochs the eclipse is not as well resolved because of its short duration in phase and hence the detection efficiency declines strongly at $N<200$. The overall detection efficiency for $N>50$ is 0.670, suggesting that there might be eclipsing sources with similar light curve properties that have not been detected in this study.

\paragraph*{ZTF J163743.49+491740.4} (J1637+4917) is the shortest period binary found in this study. Its BLS-period is 61.5\,min and the best-fit triweight parameters are 0.023 and 0.62\,mag for the eclipse duration and depth respectively (Figure~\ref{fig:eclipse1}).  Similar to J1356+5705, the eclipse is short and sharp, lasting only 85\,s.  This is reflected in the simulations too, where J1637+4917 is found to have a low detection efficiency of 0.591 at $N>50$. At $N>400$, however, J1637+4917 is nearly always detected. In the H-R diagram J1637+4917 is located in the white dwarf sequence at $G_\mathrm{BP} - G_\mathrm{RP} = 0.27$\,mag at a faint absolute magnitude of $G_\mathrm{abs}=12.65$. This gives it a probability $P_\mathrm{WD} = 1.00$ of being a white dwarf. This binary was independently discovered by the ZTF team as well and their follow-up observations confirmed it as an AM\,CVn system \citep{vanRoestel21_5amcvns}. The relatively red and faint position in the H-R diagram is quite unusual compared to known AM\,CVn systems \citep{Ramsay18}, and it indicates a very faint and evolved companion star.

\paragraph*{ZTF J021919.78+462345.1} (J0219+4623) has a period of 83.98\,min and the triweight fit yields a duration of 0.112 and a depth of 1.47\,mag (Figure~\ref{fig:eclipse1}). Its median magnitudes are $g=19.9$\,mag and $r=19.7$\,mag and the phase folded light curve features a slight asymmetry. In the H-R diagram J0219+4623 lies at $G_\mathrm{BP} - G_\mathrm{RP} = 0.47$\,mag and $G_\mathrm{abs}=10.87$\,mag, which is in a region populated by main sequence companion binaries. It therefore has a small probability $P_\mathrm{WD} = 0.36$ of being a white dwarf and can be grouped with the CV or detached white dwarf--main sequence (WDMS) candidates. %
The spectrum (Figure~\ref{fig:spectra}) shows a blue continuum with double-peaked hydrogen emission lines, as well as single-peaked HeI. This confirms J0219+4623 as a CV, with the emission lines originating in the accretion disc which is viewed at high inclination.
In the simulation the detection efficiency of J0219+4623 is 0.951 for $N>50$ which suggests that binaries with similar light curve properties would have been detected with high probability in this study.

\begin{figure*}
	\centering
		\includegraphics[width=0.85\linewidth]{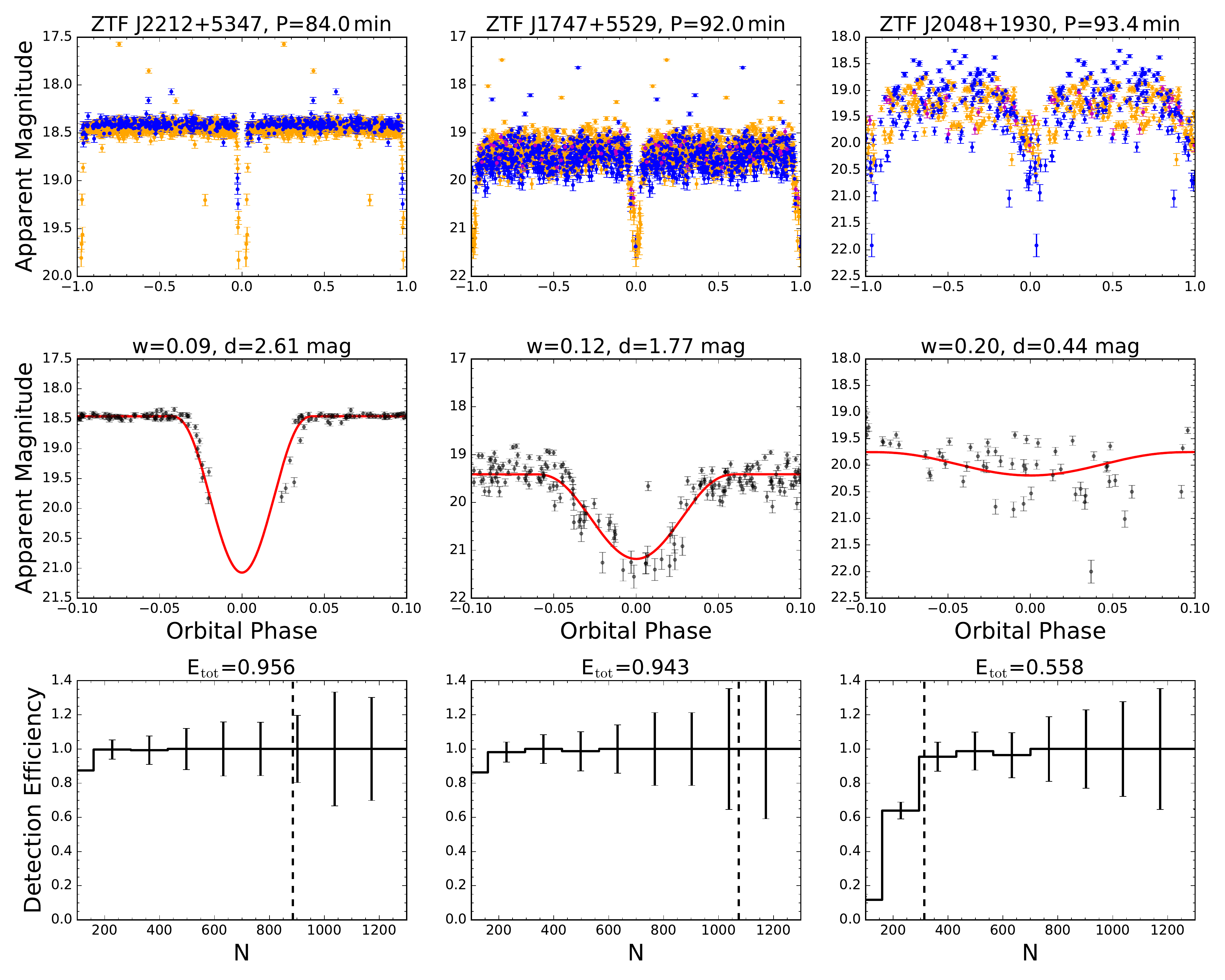}
	\caption{Same as Figure \ref{fig:eclipse1} but for different binaries. These binary sources are likely to have main sequence companions and ZTF J1747+5529 has the characteristics of a CV. The eclipse of ZTF J2212+5347 is fainter than the ZTF magnitude limit and is thus not fully resolved in depth. ZTF J2048+1930 has a noisy light curve and the light curve minimum is poorly fitted by a triweight kernel. Hence, the width and depth and the efficiency results are to be taken with a pinch of salt. We suggest that this is not an eclipsing system, but instead a magnetic cataclysmic variable.}
	\label{fig:eclipse2}
\end{figure*}

\paragraph*{ZTF J221226.96+534750.7} (J2212+5347) is one of the closest binaries in our sample, at 193\,pc \citep{bj18}. Its phase folded light curve (Figure~\ref{fig:eclipse2}) shows a deep eclipse on a period of 84.01\,min. The eclipse depth is greater than 1.5\,mag, but it is not well determined, as the object is fainter than the ZTF magnitude limit at mid-eclipse. The phase duration of the eclipse is measured to be 0.088, corresponding to 7.4\,min. On the H-R diagram, J2212+5347 lies about a magnitude above the white dwarf sequence at $G_\mathrm{abs}=11.85$ and $G_\mathrm{BP} - G_\mathrm{RP} = 0.41$\,mag, towards the binaries with dwarf main sequence companions. The long duration of the eclipse is a further suggestion that the companion is a main sequence star rather than a white dwarf. The probability of the source being a white dwarf is $P_\mathrm{WD} = 0.88$. There are two ZTF epochs in the $g$-band that lie more than $3\sigma$ above the median brightness of 18.4\,mag. These might be due to flaring on the companion star. The results of the simulation are ambiguous because of the unresolved eclipse depth. Given the relatively long eclipse duration and the the lower limit of the eclipse duration, however, the detection efficiency may be assumed to be close to 1.0. %
The spectrum (Figure~\ref{fig:spectra}) has Balmer absorption lines on a blue continuum, characteristic of a DA (hydrogen) white dwarf. We detect no obvious features of the companion above the noise of the spectrum, so the companion is much fainter than the white dwarf, even at the red end of the spectrum.

\paragraph*{ZTF J174747.26+552917.8} (J1747+5529) is a cataclysmic variable with a period of 91.99\,min. Its primary eclipse has a depth of 1.77\,mag and a phase duration of 0.125 (Figure~\ref{fig:eclipse2}). There is a characteristic rise is the light curve just before the start of the eclipse, which results from the bright spot on the edge of the accretion disc. J1747+5529 was previously identified as a CV candidate by the Catalina Real-time Transient Survey (CRTS) due to a 3.5\,mag outburst observed in 2012, but its orbital period was not known and it was not known to be eclipsing. The strong double-peaked H and HeI lines in our spectrum confirms this classification (Figure~\ref{fig:spectra}). As expected for cataclysmic variables, which have additional light from the companion star as well as the accretion disc, the probability of being a white dwarf is low, $P_\mathrm{WD} = 0.15$. Outside of the eclipse there are non-periodic brightness fluctuations of $\sim1$\,mag, which also suggest the presence of accretion. With $G_\mathrm{BP} - G_\mathrm{RP} = 0.58$\,mag and $G_\mathrm{abs}=11.09$, J1747+5529 is the reddest object in our sample. It performs well in the detection efficiency simulation, having an overall detection efficiency of 0.943 at $N>50$.

\paragraph*{ZTF J204844.69+193019.4} (J2048+1930) has a period of 93.40\,min and is characterised by strong non-periodic scatter and light curve maxima at phase $\pm0.5$ (Figure~\ref{fig:eclipse2}). The median magnitude ($19.2$\,mag) is similar in all three bands. In the H-R diagram J2048+1930 lies at $G_\mathrm{BP} - G_\mathrm{RP} = 0.16$\,mag and $G_\mathrm{abs}=8.57$ which is well above the white dwarf sequence. Its probability of being a white dwarf is $P_\mathrm{WD} = 0.48$. 

Even though J2048+1930 was identified as a potential eclipser in our search, it is not clear that the dip in the light curve really is due to an eclipse. The light curve is reminiscent of rotational variables, where the variability is caused by a bright spot rotating into and out of view. Considering the orbital period of this binary and the position on the H-R diagram close to the white dwarf--main sequence binaries, we classified this binary as a likely polar (a magnetic cataclysmic variable). The hot spot on the white dwarf results from the shock region where the material from the main sequence star is accreted onto the white dwarf along the magnetic field lines. See, for example, the folded light curve of the known low-state polar EF\,Eri \citep{Szkody10} which looks very similar to this one. To confirm this classification, we obtained the spectrum shown in Figure~\ref{fig:spectra}. As commonly seen in CVs, it has H and He in emission, but the most notable feature of the spectrum is a broad cyclotron hump at $\sim5400$\AA, which confirms J2048+1930 as a polar. The cyclotron line will vary in strength as the binary rotates, which gives rise to the variability seen in the lightcurve as well. We attribute the apparent lines at the red end of the spectrum to poorly subtracted night sky emission.
The triweight fit to the lightcurve is poor, due to the broad minimum, and as a result, the detection efficiency of 0.558 for $N>50$ is probably underestimated. J2048+1930 is included in VSX as a novalike variable MGAB-V539, with an orbital period matching ours, but we argue that neither the period nor the position in the H-R diagram are typical of novalike (high accretion rate) cataclysmic variables. 

\begin{figure*}
	\centering
		\includegraphics[width=0.85\linewidth]{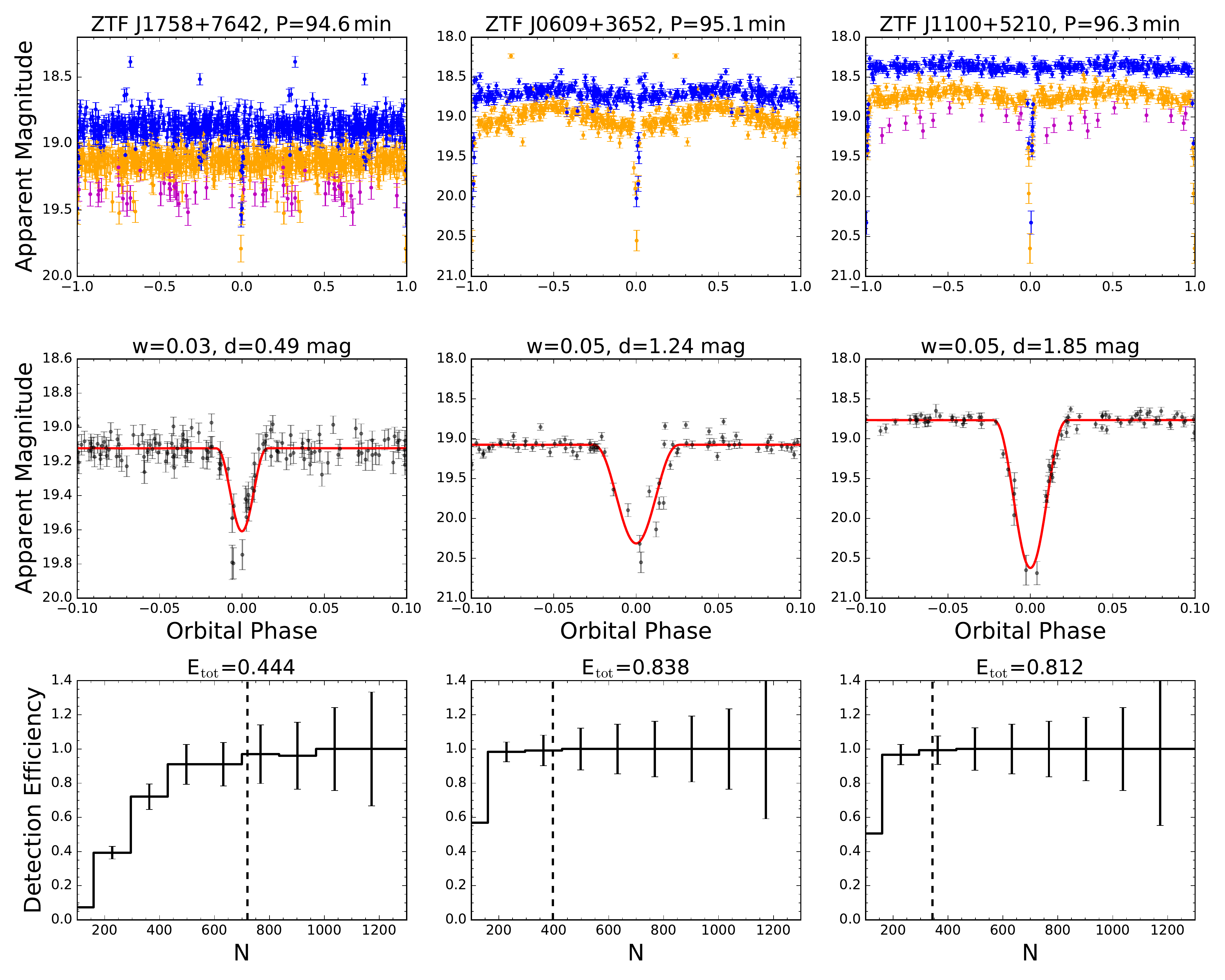}
	\caption{Same as Figure \ref{fig:eclipse1} but for different binaries. The light curve of ZTF J1758+7642 resembles the two shorter period binaries in Figure \ref{fig:eclipse1} and has a poor detection efficiency due to its shallow depth. The other two, ZTF J0609+3652 and ZTF J1100+5210, are characterised by pronounced irradiation patterns. Interestingly, ZTF\,J1100+5210 has an SDSS spectrum which does not show any signs of a companion star. Together with its light curve features this suggest that the companion might be a brown dwarf.}
	\label{fig:eclipse3}
\end{figure*}

\paragraph*{ZTF J175812.85+764216.8} (J1758+7642) is a 94.56\,min binary with a short eclipse duration of 0.033 in phase (3.12 min) and a depth of 0.49\,mag. The light curve has median magnitudes of $g=18.9$\,mag, $r=19.1$\,mag and $i=19.4$\,mag and appears to be flat out of the eclipse (Figure~\ref{fig:eclipse3}). In the H-R diagram J1758+7642 is located on the blue side of the white dwarf sequence with $G_\mathrm{BP}-G_\mathrm{RP} \approx -0.18$ and $G_\mathrm{abs} = 10.25$. Its position on the H-R diagram and probability $P_\mathrm{WD} = 1.00$ of being a white dwarf, makes this a strong candidate for an new eclipsing double degenerate or ELM binary. %
Its spectrum is dominated by a DA white dwarf and no features from a companion is visible. Radial velocity studies will be needed to measure the masses of the component stars. Due to its short eclipse duration and shallow depth, J1758+7642 has a low detection efficiency of 0.44 for $N>50$. %

\paragraph*{ZTF J060914.37+365202.9} (J0609+3652) has a pronounced irradiation pattern with a period of 95.10\,min (Figure~\ref{fig:eclipse3}). The eclipse duration and depth found with the triweight fit are 0.055 in phase and 1.24\,mag respectively. The light curve has median magnitudes $g=18.7$\,mag and $r=19.0$\,mag and the modulation pattern has a higher amplitude in $r$ than in $g$, which is characteristic of irradiation. 
On the H-R diagram J0609+3652 lies above the white dwarf sequence at $G_\mathrm{abs} = 8.87$ and $G_\mathrm{BP}-G_\mathrm{RP} = -0.06$, among the white dwarf--main sequence binaries. Its probability of being a white dwarf is $P_\mathrm{WD} = 0.90$. In the simulation, J0609+3652 was detected with an efficiency of 0.838 for $N>50$. This star was also reported to VSX by amateur astronomer Gabriel Murawski. Its orbital period is listed as uncertain, but agrees with the period we derive here. No spectrum is available for this system yet.

\begin{figure}
	\centering
	\includegraphics[width=\columnwidth]{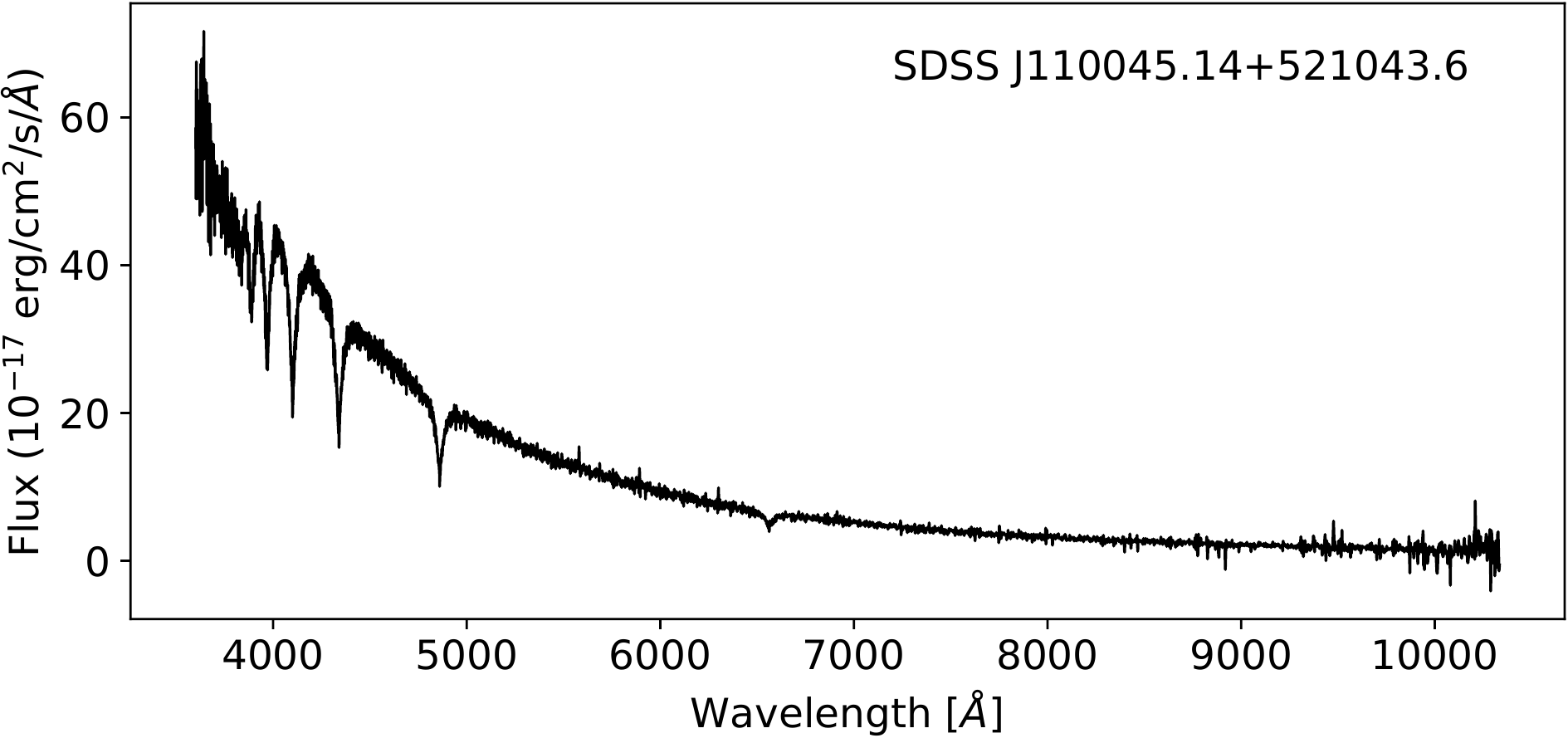}~\\\vspace{2mm}
	\includegraphics[width=\columnwidth]{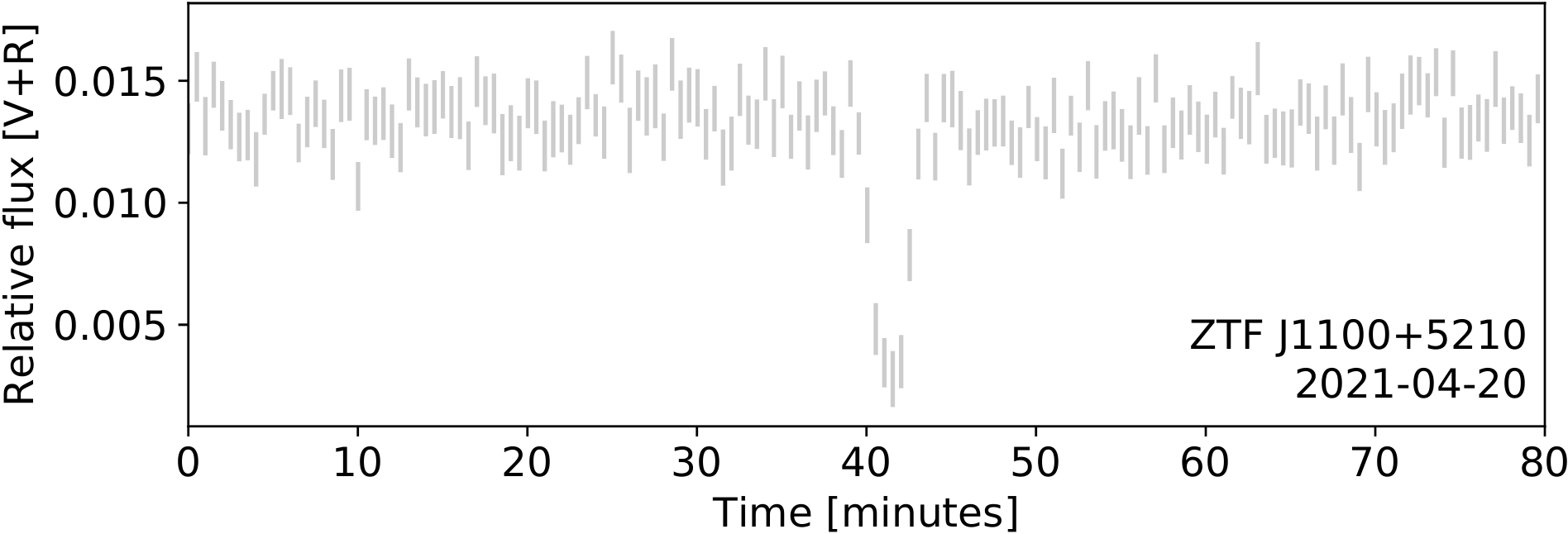}
	\caption{{\em Top panel:} The SDSS spectrum of ZTF/SDSS J110045.14+521043.6. The spectrum is that of a hydrogen white dwarf, with no features from a companion star seen. {\em Bottom panel:} $V+R$-band light curve of ZTF\,J110045.14+521043.6, obtained at the Liverpool Telescope.
	\label{fig:wd1100} }
\end{figure}

\paragraph*{ZTF J110045.14+521043.6} (J1100+5210) has a period of 96.31\,min, an eclipse duration of 0.047 in phase (4.53\,min) and a depth of 1.85\,mag (Figure~\ref{fig:eclipse3}). Its light curve has an irradiation pattern with an amplitude slightly larger in the $r$-band than in the $g$-band. J1100+5210 is located at the blue end of the white dwarf sequence in the H-R diagram, at $G_\mathrm{BP}-G_\mathrm{RP} = -0.21$ and $G_\mathrm{abs} = 9.94$. Based on the analysis of \citet{gentile2019}, its probability of being a white dwarf is $P_\mathrm{WD}=0.99$. A spectrum of this star is available in SDSS (Fig.~\ref{fig:wd1100}), and it shows it to be a $23\,886\pm430$~K hydrogen white dwarf \citep{Anguiano17}. No features from the companion star are visible in the spectrum. Together with the irradiation in the light curve and the length of the eclipse, this suggests that the companion might be a brown dwarf. The object is not detected in the 2~micron all-sky survey (2MASS) but there are faint detections at this position in the W1 and W2 bands (3.4 and 4.6$\mu$m) of the CatWISE mid-infrared survey \citep{catwise}. Infrared spectroscopy will be needed to determine the spectral type of the donor star in this binary. If confirmed, this will only be the fifth example of an eclipsing white dwarf--brown dwarf binary \citep[see e.g.][]{Casewell20,vanRoestel21_WDBD}.
In the simulation it is detected with an efficiency of 0.81, suggesting that sources with similar light curves would have probably been detected in this study. 
We confirmed the eclipse, and hence the binary nature of the object, at the Liverpool Telescope on 2021 April 20 (Figure~\ref{fig:wd1100}). We used the RISE camera with its broad $V+R$ filter and observed the target for a total of 100\,min, using 30\,s exposures. Conditions were clear, with a 58\% illuminated moon and subarcsecond seeing. The eclipse is clearly detected, but given the broad filter of the camera, we don't detect the irradiation seen in the ZTF-$r$ and $i$ band light curves. Future high time resolution observations will allow us to model the light curve and measure its binary parameters. 

\begin{figure*}
	\centering
		\includegraphics[width=0.85\linewidth]{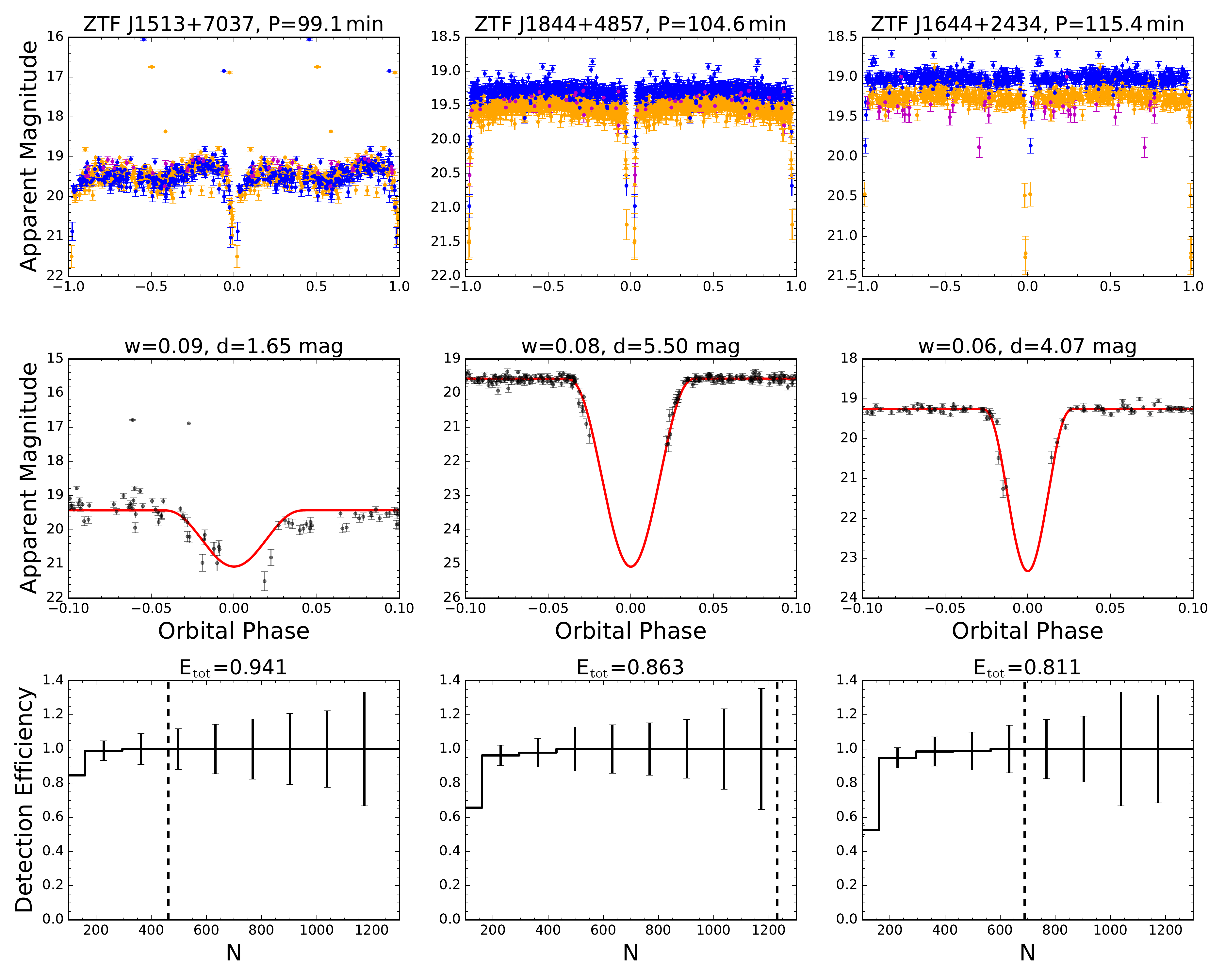}
	\caption{Same as Figure \ref{fig:eclipse1} but for different binaries. The sources in this figure probably all have main sequence companions. ZTF J1513+7037 is a clear CV candidate and the only one among the newly discovered ones to have a clearly assymmetric light curve. ZTF J1844+4857 and ZTF J1644+2435 have light curves that are similar in appearance, both having a slight irradiation pattern and eclipses that are not resolved in depth due to the ZTF magnitude limit.}
	\label{fig:eclipse4}
\end{figure*}

\paragraph*{ZTF J151332.97+703722.4} (J1513+7037) is a CV with a period of 99.11\,min. Its primary eclipse has a duration of 0.088 in phase and a depth of 1.65\,mag (Figure~\ref{fig:eclipse4}). It was also detected by the \textit{Gaia}\, Science Alerts system \citep{hodgkin21} as Gaia17bcx, following the detection of a 2~mag outburst, but its period was not known before nor that it was eclipsing. The ZTF light curve includes a number of bright detections due to similar outbursts. It also features ellipsoidal modulation and a pronounced asymmetry of the eclipse. The ingress is steeper than the egress and there is a rise just before the start of the eclipse. These features are commonly seen in the light curves of eclipsing CVs, due to the various components (white dwarf, bright spot, accretion disc) entering and leaving the eclipse \citep[see e.g.][]{mcallister19}. All features of this source are therefore strongly in favour of it being a CV. %
The seeing was quite poor when we took the spectrum of this system and hence it suffered significant slit losses. The continuum is therefore faint and its shape not well defined (Figure~\ref{fig:spectra}), but the spectrum has strong double-peaked H$\alpha$, H$\beta$ and H$\gamma$ emission, characteristic of a high-inclination CV. %
On the H-R diagram J1513+7037 can be found at $G_\mathrm{BP}-G_\mathrm{RP} = 0.50$ and $G_\mathrm{abs} = 10.99$. It has a low probability $P_\mathrm{WD} = 0.27$ of being a white dwarf, as expected for a binary with a main sequence companion. J1513+7037 has a high detection efficiency of 0.94 for $N>50$.

\paragraph*{ZTF J184434.39+485736.5} (J1844+4857) has a period of 104.64\,min, an eclipse duration of 0.080 in phase and a depth greater than 2\,mag. The depth is unresolved because of the ZTF magnitude limit and hence the triweight depth may be overestimated. The light curve shows an irradiation pattern that is most noticeable in the $r$-band (Figure~\ref{fig:eclipse4}).
On the H-R diagram, J1844+4857 lies in a region between the white dwarf sequence and the main sequence at $G_\mathrm{BP}-G_\mathrm{RP} = 0.18$ and $G_\mathrm{abs} = 9.83$, with a probability of being a white dwarf $P_\mathrm{WD} = 0.77$. J1844+4857 is likely a white dwarf with a main sequence companion, but the nature of the companion could not be identified from the spectrum, as it is completely dominated by the white dwarf. %
Similar to J2212+5347, the efficiency of 0.86 at $N>50$ might be overestimated due to the unresolved eclipse depth. This binary was also identified by Gabriel Murawski, who listed it on VSX as  MGAB-V1137.

\paragraph*{ZTF J164441.19+243428.2} (J1644+2434) has an irradiation pattern with a period of 115.35\,min (Figure~\ref{fig:eclipse4}). The eclipse duration and depth found by fitting a triweight kernel are 0.056 in phase and 4.07\,mag respectively. The eclipse minimum exceeds the ZTF magnitude limit so the triweight estimate of the depth may be overestimated. On the H-R diagram J1644+2434 lies about a magnitude above the white dwarf sequence, at $G_\mathrm{BP}-G_\mathrm{RP} = 0.19$ and $G_\mathrm{abs} = 10.97$. Its probability of being a white dwarf is $P_\mathrm{WD} = 0.94$, which suggests that the white dwarf dominates the light and that the companion is a low luminosity object. Our spectrum confirms the dominance of the white dwarf, and as expected, no features from the companion star are seen. Further radial velocity studies are needed to determine the nature of this binary. %
The detection efficiency at $N>50$ is 0.81 which might be lower due to the unresolved eclipse depth. It was previously identified as a variable white dwarf MGAB-V1139 by Gabriel Murawski, but its period was not known.

\begin{figure*}
	\centering
		\includegraphics[width=0.85\linewidth]{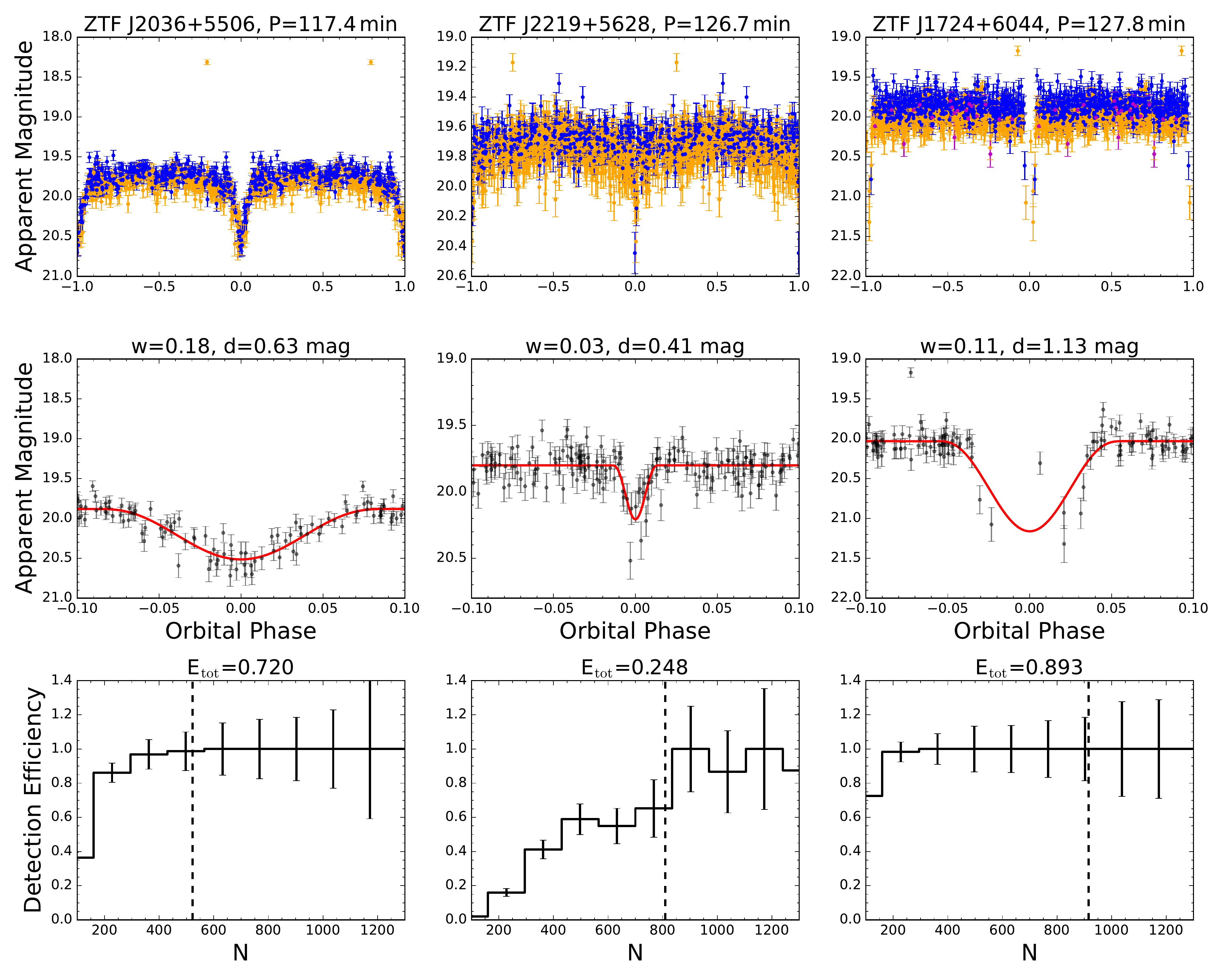}
	\caption{Same as Figure \ref{fig:eclipse1} but for different binaries. All sources in this figure probably have main sequence companions. ZTF J2036+5506 has a clearly visible and broad pimary and secondary eclipse. ZTF 2219+5628 has some kind of modulation and an extremely short and shallow eclipse and thus a poor detection efficiency. ZTF J1724+6044 has an eclipse which is possibly underresolved in depth due to the ZTF magnitude limit.}
	\label{fig:eclipse5}
\end{figure*}

\paragraph*{ZTF J203601.67+550648.7} (J2036+5506) lies above the white dwarf sequence on the H-R diagram at $G_\mathrm{BP}-G_\mathrm{RP} = -0.08$ and $G_\mathrm{abs} = 8.84$. Its \textit{Gaia}\, parallax has a large uncertainty, leading to a large uncertainty in its absolute magnitude. Nevertheless, its white dwarf probability is high, at $P_\mathrm{WD} = 0.95$. This binary is included in the ZTF catalogue of periodic variable stars \citep{chen20} as an Algol-type eclipsing system with period 0.1631190\,d, but this classification does not match its position in the H-R diagram. Our analysis also prefers half that value as the orbital period, $P=0.0815591$\,d = 117.45\,min. The reason for this preference is that the light curve, when folded on this period, is suggestive of a secondary eclipse at phases $\pm0.5$ (Figure~\ref{fig:eclipse5}). A Lomb-Scargle periodogram also selects this as the strongest period in the light curve and two harmonics of this period are also seen. We do not detect power above the noise at $P=0.1631190$\,d. The primary eclipse has a relatively long duration of 0.177 in phase and a depth of 0.63 mag. %
The Balmer lines in the spectrum are weak and narrow, and the spectrum looks more like a hot subdwarf, rather than a white dwarf. Its position in the H-R diagram appears far away from where subdwarfs are normally found, but the parallax of this system is not well constrained (parallax/error = 2). An improved parallax in future \textit{Gaia}\, data releases could move it to the region occupied by subdwarfs. Based on the spectrum, we classify this as a subdwarf--main sequence binary. %
There is one outlier in the $g$-band light curve, at $\sim1.5$\,mag brighter than the rest of the light curve, that may be due to a flare on the companion star. In the simulation J2036+5506 has an efficiency of 0.72 at $N>50$ and is detected for all light curves with $N>600$.

\paragraph*{ZTF J221912.73+562805.4} (J2219+5628) is dominated by a strong irradiation pattern with a period of 126.65\,min (Figure~\ref{fig:eclipse5}). The eclipse is short and shallow (3.5\,min, 0.41\, mag) which causes a comparatively low detection efficiency of 0.025 at $N>50$. This is probably a partial eclipse of the white dwarf. In the H-R diagram it is located well above the white dwarf sequence, at $G_\mathrm{BP}-G_\mathrm{RP} = 0.29$ and $G_\mathrm{abs} = 8.71$. This region is known to be populated by binaries with main sequence companions, which is also reflected in the low probability $P_\mathrm{WD}=0.4$ of being a white dwarf. The spectrum is again dominated by the white dwarf, with no features from a companion seen.

\paragraph*{ZTF J172400.04+604451.8} (J1724+6044) is a new eclipsing binary featuring a weak irradiation pattern and a clear, sharp eclipse on a period of 127.77\,min (Figure~\ref{fig:eclipse5}). The eclipse has a relatively long duration of 0.113 in phase (14.4\,min) and a depth of at least 1.13\,mag. The median apparent magnitudes of the light curve are quite faint at $g=19.9$\,mag and $r=20.0$\,mag, so it is below the ZTF magnitude limit at mid-eclipse.
J1724+6044 lies slightly above the white dwarf sequence on the H-R diagram, at $G_\mathrm{BP}-G_\mathrm{RP} = 0.15$ and $G_\mathrm{abs} = 10.10$. Its probability of being a white dwarf is $P_\mathrm{WD}=0.76$, with the spectrum showing a DA white dwarf with little or no H$\alpha$. Its detection efficiency is 0.89 at $N>50$, saturating at $N>400$.

\begin{figure*}
	\centering
		\includegraphics[width=0.85\linewidth]{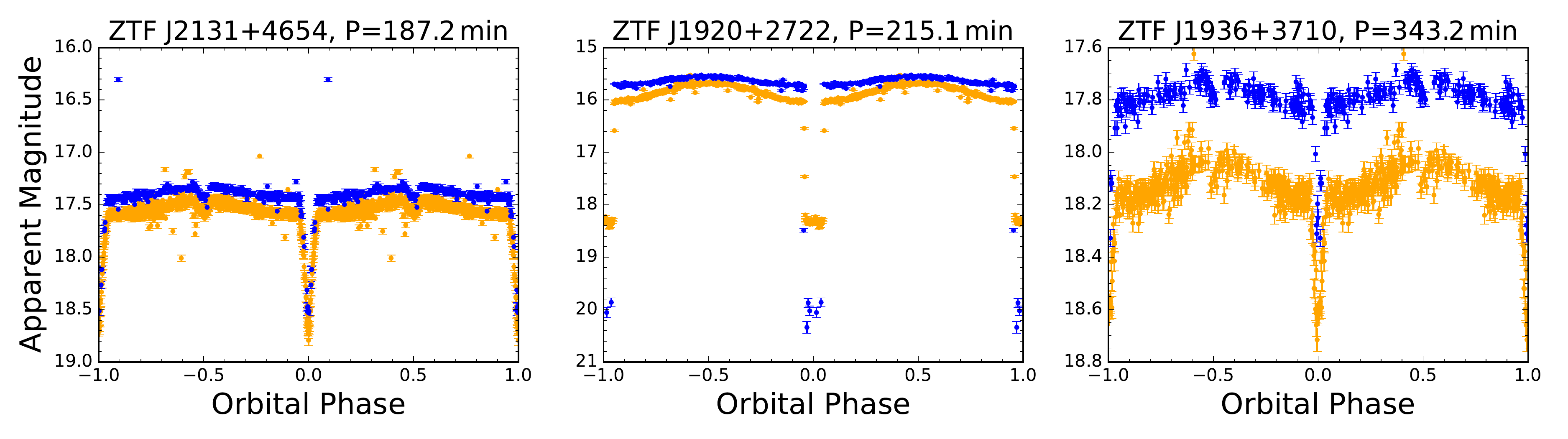}
	\caption{Similar to Figure \ref{fig:eclipse1} but for different binaries. These are the longest period binaries detected in this study. Their periods exceed the nominal search range, because they were detected at one of their harmonics. Hence, they have not been included in the efficiency simulations. All have main sequence components and ZTF J2131+4654 and ZTF J1936+3710 have visible secondary eclipses. ZTF J1920+2722 is the only source to have a total eclipse and eclipse depths that clearly differ in the $g-$ and $r-$bands.}
	\label{fig:eclipse6}
\end{figure*}

\paragraph*{ZTF J213141.44+465430.0} (J2131+4654) was initially detected as a binary candidate at a harmonic of its orbital period (Section~\ref{sec:BLS}). It has a period of 187.15\,min and a pronounced irradiation pattern with well resolved primary and secondary eclipses (Figure~\ref{fig:eclipse6}). The best fit triweight-parameters are 0.095 in phase and 1.04\,mag for the eclipse duration and depth respectively. 
J2131+4654 lies well above the white dwarf sequence, at $G_\mathrm{BP}-G_\mathrm{RP} = 0.03$ and $G_\mathrm{abs} = 6.19$ and its probability of being a white dwarf is close to zero. Hence, this binary involves a main sequence companion and a subdwarf, rather than a white dwarf (See Figure~\ref{fig:spectra}). This binary was included in the ZTF periodic variable stars catalogue of \citet{chen20} as a W\,UMa-type star with a period twice our derived period. Given the clear detection of the shallow secondary eclipse in this system, we are confident of our period, and suggest that the W\,UMa classification is incorrect. 

\paragraph*{ZTF J192014.13+272218.1} (J1920+2722) has a period of 215.09\,min and is the brightest eclipsing binary discovered in this study. Its median magnitudes are $g=15.7$\,mag and $r=15.8$\,mag. At 190\,pc \citep{bj18} it is also the closest binary in our sample. The orbital period exceeds the nominal period search range, because it was identified as a candidate at one of its harmonics. The light curve features an irradiation pattern with a higher amplitude in $r$ than in $g$ and a deep, flat primary eclipse indicative of totality (Figure~\ref{fig:eclipse6}). Unlike the other binaries, J1920+2722 has an eclipse that is significantly deeper in $g$ than in $r$. This indicates that the eclipsed star is very blue, in this case a hot white dwarf. The eclipse depth is 2.3\,mag in $r$ and exceeds 4\,mag in $g$. Unfortunately, GALEX did not cover this part of the sky so no ultraviolet measurements are available. In the H-R diagram J1920+2722 lies at $G_\mathrm{BP}-G_\mathrm{RP} = 0.37$ and $G_\mathrm{abs} = 9.16$, well above the white dwarf sequence. Its probability of being a white dwarf is thus quite low, $P_\mathrm{WD}=0.22$. The red end of the spectrum shows the TiO absorption bands from the M-dwarf companion, along with the hydrogen white dwarf which dominates at blue wavelengths. This confirms J1920+2722 as a WDMS binary. This binary was also identified as an eclipsing binary by Gabriel Murawski, and submitted to VSX as MGAB-V287.

\paragraph*{ZTF J193604.88+371017.2} (J1936+3710) is the longest period binary found in this study. It has a period of 343.23\,min and it was initially detected at a harmonic of this period. The light curve has a pronounced irradiation pattern and both the primary and the secondary eclipses are resolved (Figure~\ref{fig:eclipse6}). The eclipse duration and depth of the primary found by fitting a triweight kernel are 0.095 in phase and 0.45\,mag respectively.
J1936+3710 is one of the bluest targets in our sample ($G_\mathrm{BP}-G_\mathrm{RP} = -0.17$) and it lies well above the white dwarf sequence at $G_\mathrm{abs} = 5.91$. It has a low probability $P_\mathrm{WD} = 0.19$ of being a white dwarf. Its location on the H-R diagram suggest that this might be a subdwarf rather than a white dwarf binary and our spectrum in Figure~\ref{fig:spectra} confirms that.

\section{Summary}
\label{sec:summary}

We presented the first results from our ongoing search for eclipsing white dwarf binaries using \textit{Gaia} and the ZTF. ZTF provides high quality light curves for a large number of stars, while the astrometric and photometric data from \textit{Gaia} allows us to identify the nature of these binaries from their position on the H-R diagram. Note that although the \textit{Gaia} DR2 parameters were used in this work, the Early Data Release 3 (EDR3) sourceID is identical for all our sources, and the differences in parallax and colour are small. 

We focused our search on short periods, 5.7--144\,min, and identified 18 new binaries, although three of these were detected at a harmonic of their true orbital period and have periods longer than the nominal search range.  
Within the 18 new binary systems reported here, we identify two binaries which potentially contain ELM white dwarfs, one confirmed AM\,CVn star, one candidate white dwarf--brown dwarf binary, four confirmed cataclysmic variables, seven detached white dwarf--main sequence binaries, and three detached subdwarf--main sequence binaries. Ultimately, although certain classes of binaries can be ruled out based on the light curves and their positions in the H-R diagram, we found that spectroscopy is needed for a robust classification, because of the variety of systems in this part of the H-R diagram. Our spectroscopic follow-up confirmed the majority of these classifications, but some uncertainty remain for the WDMS binaries where the companions were not directly detected in our spectra.

The binaries detected so far are at large heliocentric distances (compared to single white dwarfs) and imply significant heights above the Galactic disc.

Although not all known eclipsers are detectable in the ZTF data, a crude estimate of the completeness of our search suggest that more white dwarf binaries remain to be found in the ZTF footprint, with a slight adjustment of our light curve selection criteria. We also find that the number of epochs in the light curve is a key factor in our ability to detect eclipsing systems, especially for eclipses that are shallow or have a short phase duration. Future searches could also include more general variability, not just eclipses, to reveal white dwarf binaries \citep[cf.][]{Guidry20} as well as longer orbital periods.

\section*{Acknowledgements}

Based on observations obtained with the Samuel Oschin 48-inch Telescope at the Palomar Observatory as part of the Zwicky Transient Facility project. ZTF is supported by the National Science Foundation under Grant No. AST-1440341 and a collaboration including Caltech, IPAC, the Weizmann Institute for Science, the Oskar Klein Center at Stockholm University, the University of Maryland, the University of Washington, Deutsches Elektronen-Synchrotron and Humboldt University, Los Alamos National Laboratories, the TANGO Consortium of Taiwan, the University of Wisconsin at Milwaukee, and Lawrence Berkeley National Laboratories. Operations are conducted by COO, IPAC, and UW. 

This work also includes data from the European Space Agency (ESA) mission \textit{Gaia}\, (\url{https://www.cosmos.esa.int/gaia}), processed by the \textit{Gaia}\, Data Processing and Analysis Consortium (DPAC,
\url{https://www.cosmos.esa.int/web/gaia/dpac/consortium}). Funding for the DPAC has been provided by national institutions, in particular the institutions participating in the \textit{Gaia}\, Multilateral Agreement. Further details of funding authorities and individuals contributing to the success of the mission is shown at \url{https://gea.esac.esa.int/ archive/documentation/GDR2/Miscellaneous/sec\_acknowl/}.

This paper made use of the Whole Sky Database (WSDB) created by Sergey Koposov and maintained at the Institute of Astronomy, Cambridge with financial support from the Science and Technology Facilities Council (STFC) and the European Research Council (ERC).

PMK was funded by the Swiss Study Foundation and by an Isaac Newton studentship.
EB and STH are funded by the Science and Technology Facilities Council grant ST/S000623/1.

We thank Ingrid Pelisoli, Joris Vos and Alina Istrate for providing the ELM tracks shown in Figure~\ref{fig:HR_eclipses}, and an anonymous referee whose comments helped to improve the clarity of this paper.

\section*{Data Availability}
All \textit{Gaia} and ZTF data used in this paper are publicly available. Details on how to access the data can be found on their websites: \url{https://www.cosmos.esa.int/web/gaia/data-release-2} and \url{https://www.ztf.caltech.edu/page/dr3}. The spectra will be made available upon reasonable request to the corresponding authors.

\bibliographystyle{mnras}
\bibliography{references.bib}

\begin{thebibliography}{}
\makeatletter
\relax
\def\mn@urlcharsother{\let\do\@makeother \do\$\do\&\do\#\do\^\do\_\do\%\do\~}
\def\mn@doi{\begingroup\mn@urlcharsother \@ifnextchar [ {\mn@doi@}
  {\mn@doi@[]}}
\def\mn@doi@[#1]#2{\def\@tempa{#1}\ifx\@tempa\@empty \href
  {http://dx.doi.org/#2} {doi:#2}\else \href {http://dx.doi.org/#2} {#1}\fi
  \endgroup}
\def\mn@eprint#1#2{\mn@eprint@#1:#2::\@nil}
\def\mn@eprint@arXiv#1{\href {http://arxiv.org/abs/#1} {{\tt arXiv:#1}}}
\def\mn@eprint@dblp#1{\href {http://dblp.uni-trier.de/rec/bibtex/#1.xml}
  {dblp:#1}}
\def\mn@eprint@#1:#2:#3:#4\@nil{\def\@tempa {#1}\def\@tempb {#2}\def\@tempc
  {#3}\ifx \@tempc \@empty \let \@tempc \@tempb \let \@tempb \@tempa \fi \ifx
  \@tempb \@empty \def\@tempb {arXiv}\fi \@ifundefined
  {mn@eprint@\@tempb}{\@tempb:\@tempc}{\expandafter \expandafter \csname
  mn@eprint@\@tempb\endcsname \expandafter{\@tempc}}}

\bibitem[\protect\citeauthoryear{{Abolfathi} et~al.,}{{Abolfathi}
  et~al.}{2018}]{abolfathi2018}
{Abolfathi} B.,  et~al., 2018, \mn@doi [\apjs] {10.3847/1538-4365/aa9e8a},
  \href {https://ui.adsabs.harvard.edu/abs/2018ApJS..235...42A} {235, 42}

\bibitem[\protect\citeauthoryear{{Anguiano}, {Rebassa-Mansergas},
  {Garc{\'\i}a-Berro}, {Torres}, {Freeman}  \& {Zwitter}}{{Anguiano}
  et~al.}{2017}]{Anguiano17}
{Anguiano} B.,  {Rebassa-Mansergas} A.,  {Garc{\'\i}a-Berro} E.,  {Torres} S.,
  {Freeman} K.~C.,   {Zwitter} T.,  2017, \mn@doi [\mnras]
  {10.1093/mnras/stx796}, \href
  {https://ui.adsabs.harvard.edu/abs/2017MNRAS.469.2102A} {469, 2102}

\bibitem[\protect\citeauthoryear{{Astropy Collaboration} et~al.,}{{Astropy
  Collaboration} et~al.}{2013}]{astropy2013}
{Astropy Collaboration} et~al., 2013, \mn@doi [\aap]
  {10.1051/0004-6361/201322068}, \href
  {https://ui.adsabs.harvard.edu/abs/2013A&A...558A..33A} {558, A33}

\bibitem[\protect\citeauthoryear{{Astropy Collaboration} et~al.,}{{Astropy
  Collaboration} et~al.}{2018}]{astropy2018}
{Astropy Collaboration} et~al., 2018, \mn@doi [\aj] {10.3847/1538-3881/aabc4f},
  \href {https://ui.adsabs.harvard.edu/abs/2018AJ....156..123A} {156, 123}

\bibitem[\protect\citeauthoryear{{Bailer-Jones}, {Rybizki}, {Fouesneau},
  {Mantelet}  \& {Andrae}}{{Bailer-Jones} et~al.}{2018}]{bj18}
{Bailer-Jones} C.~A.~L.,  {Rybizki} J.,  {Fouesneau} M.,  {Mantelet} G.,
  {Andrae} R.,  2018, \mn@doi [\aj] {10.3847/1538-3881/aacb21}, \href
  {https://ui.adsabs.harvard.edu/abs/2018AJ....156...58B} {156, 58}

\bibitem[\protect\citeauthoryear{{Bellm} et~al.,}{{Bellm}
  et~al.}{2019}]{bellm2018}
{Bellm} E.~C.,  et~al., 2019, \mn@doi [\pasp] {10.1088/1538-3873/aaecbe}, \href
  {https://ui.adsabs.harvard.edu/abs/2019PASP..131a8002B} {131, 018002}

\bibitem[\protect\citeauthoryear{{Belokurov} et~al.,}{{Belokurov}
  et~al.}{2020}]{Belokurov2020}
{Belokurov} V.,  et~al., 2020, \mn@doi [\mnras] {10.1093/mnras/staa1522}, \href
  {https://ui.adsabs.harvard.edu/abs/2020MNRAS.496.1922B} {496, 1922}

\bibitem[\protect\citeauthoryear{{Bergeron}, {Dufour}, {Fontaine}, {Coutu},
  {Blouin}, {Genest-Beaulieu}, {B{\'e}dard}  \& {Rolland}}{{Bergeron}
  et~al.}{2019}]{Bergeron19}
{Bergeron} P.,  {Dufour} P.,  {Fontaine} G.,  {Coutu} S.,  {Blouin} S.,
  {Genest-Beaulieu} C.,  {B{\'e}dard} A.,   {Rolland} B.,  2019, \mn@doi [\apj]
  {10.3847/1538-4357/ab153a}, \href
  {https://ui.adsabs.harvard.edu/abs/2019ApJ...876...67B} {876, 67}

\bibitem[\protect\citeauthoryear{{Brown} et~al.,}{{Brown}
  et~al.}{2020}]{brown2020elm}
{Brown} W.~R.,  et~al., 2020, \mn@doi [\apj] {10.3847/1538-4357/ab63cd}, \href
  {https://ui.adsabs.harvard.edu/abs/2020ApJ...889...49B} {889, 49}

\bibitem[\protect\citeauthoryear{{Burdge} et~al.,}{{Burdge}
  et~al.}{2019}]{burdge2019}
{Burdge} K.~B.,  et~al., 2019, \mn@doi [\nat] {10.1038/s41586-019-1403-0},
  \href {https://ui.adsabs.harvard.edu/abs/2019Natur.571..528B} {571, 528}

\bibitem[\protect\citeauthoryear{{Burdge} et~al.,}{{Burdge}
  et~al.}{2020a}]{Burdge20}
{Burdge} K.~B.,  et~al., 2020a, \mn@doi [\apj] {10.3847/1538-4357/abc261},
  \href {https://ui.adsabs.harvard.edu/abs/2020ApJ...905...32B} {905, 32}

\bibitem[\protect\citeauthoryear{{Burdge} et~al.,}{{Burdge}
  et~al.}{2020b}]{Burdge20_9min}
{Burdge} K.~B.,  et~al., 2020b, \mn@doi [\apjl] {10.3847/2041-8213/abca91},
  \href {https://ui.adsabs.harvard.edu/abs/2020ApJ...905L...7B} {905, L7}

\bibitem[\protect\citeauthoryear{{Campbell} et~al.,}{{Campbell}
  et~al.}{2015}]{Campbell15}
{Campbell} H.~C.,  et~al., 2015, \mn@doi [\mnras] {10.1093/mnras/stv1224},
  \href {https://ui.adsabs.harvard.edu/abs/2015MNRAS.452.1060C} {452, 1060}

\bibitem[\protect\citeauthoryear{{Casewell} et~al.,}{{Casewell}
  et~al.}{2020}]{Casewell20}
{Casewell} S.~L.,  et~al., 2020, \mn@doi [\mnras] {10.1093/mnras/staa1608},
  \href {https://ui.adsabs.harvard.edu/abs/2020MNRAS.497.3571C} {497, 3571}

\bibitem[\protect\citeauthoryear{{Chen}, {Wang}, {Deng}, {de Grijs}, {Yang}  \&
  {Tian}}{{Chen} et~al.}{2020}]{chen20}
{Chen} X.,  {Wang} S.,  {Deng} L.,  {de Grijs} R.,  {Yang} M.,   {Tian} H.,
  2020, \mn@doi [\apjs] {10.3847/1538-4365/ab9cae}, \href
  {https://ui.adsabs.harvard.edu/abs/2020ApJS..249...18C} {249, 18}

\bibitem[\protect\citeauthoryear{{Coughlin} et~al.,}{{Coughlin}
  et~al.}{2020}]{coughlin2020}
{Coughlin} M.~W.,  et~al., 2020, \mn@doi [\mnras] {10.1093/mnrasl/slaa044},
  \href {https://ui.adsabs.harvard.edu/abs/2020MNRAS.494L..91C} {494, L91}

\bibitem[\protect\citeauthoryear{{Dhillon} et~al.,}{{Dhillon}
  et~al.}{2014}]{Dhillon14}
{Dhillon} V.~S.,  et~al., 2014, \mn@doi [\mnras] {10.1093/mnras/stu1660}, \href
  {https://ui.adsabs.harvard.edu/abs/2014MNRAS.444.4009D} {444, 4009}

\bibitem[\protect\citeauthoryear{{Eisenhardt} et~al.,}{{Eisenhardt}
  et~al.}{2020}]{catwise}
{Eisenhardt} P. R.~M.,  et~al., 2020, \mn@doi [\apjs]
  {10.3847/1538-4365/ab7f2a}, \href
  {https://ui.adsabs.harvard.edu/abs/2020ApJS..247...69E} {247, 69}

\bibitem[\protect\citeauthoryear{{Gaia Collaboration} et~al.,}{{Gaia
  Collaboration} et~al.}{2018}]{gaia2018}
{Gaia Collaboration} et~al., 2018, \mn@doi [\aap]
  {10.1051/0004-6361/201833051}, \href
  {https://ui.adsabs.harvard.edu/abs/2018A&A...616A...1G} {616, A1}

\bibitem[\protect\citeauthoryear{{Gaia Collaboration} et~al.,}{{Gaia
  Collaboration} et~al.}{2019}]{Eyer19}
{Gaia Collaboration} et~al., 2019, \mn@doi [\aap]
  {10.1051/0004-6361/201833304}, \href
  {https://ui.adsabs.harvard.edu/abs/2019A&A...623A.110G} {623, A110}

\bibitem[\protect\citeauthoryear{{Gentile Fusillo} et~al.,}{{Gentile Fusillo}
  et~al.}{2019}]{gentile2019}
{Gentile Fusillo} N.~P.,  et~al., 2019, \mn@doi [\mnras]
  {10.1093/mnras/sty3016}, \href
  {https://ui.adsabs.harvard.edu/abs/2019MNRAS.482.4570G} {482, 4570}

\bibitem[\protect\citeauthoryear{{Green} et~al.,}{{Green}
  et~al.}{2018}]{Green18}
{Green} M.~J.,  et~al., 2018, \mn@doi [\mnras] {10.1093/mnras/sty299}, \href
  {https://ui.adsabs.harvard.edu/abs/2018MNRAS.476.1663G} {476, 1663}

\bibitem[\protect\citeauthoryear{{Guidry} et~al.,}{{Guidry}
  et~al.}{2021}]{Guidry20}
{Guidry} J.~A.,  et~al., 2021, \mn@doi [\apj] {10.3847/1538-4357/abee68}, \href
  {https://ui.adsabs.harvard.edu/abs/2021ApJ...912..125G} {912, 125}

\bibitem[\protect\citeauthoryear{{Hardy} et~al.,}{{Hardy}
  et~al.}{2017}]{Hardy17}
{Hardy} L.~K.,  et~al., 2017, \mn@doi [\mnras] {10.1093/mnras/stw3051}, \href
  {https://ui.adsabs.harvard.edu/abs/2017MNRAS.465.4968H} {465, 4968}

\bibitem[\protect\citeauthoryear{{Hodgkin} et~al.}{{Hodgkin}
  et~al.}{2021}]{hodgkin21}
{Hodgkin} S.~T.,  et~al., 2021, Astronomy \& Astrophysics

\bibitem[\protect\citeauthoryear{{Inight}, {G{\"a}nsicke}, {Breedt}, {Marsh},
  {Pala}  \& {Raddi}}{{Inight} et~al.}{2021}]{Inight21}
{Inight} K.,  {G{\"a}nsicke} B.~T.,  {Breedt} E.,  {Marsh} T.~R.,  {Pala}
  A.~F.,   {Raddi} R.,  2021, \mn@doi [\mnras] {10.1093/mnras/stab753}, \href
  {https://ui.adsabs.harvard.edu/abs/2021MNRAS.504.2420I} {504, 2420}

\bibitem[\protect\citeauthoryear{{Istrate}, {Marchant}, {Tauris}, {Langer},
  {Stancliffe}  \& {Grassitelli}}{{Istrate} et~al.}{2016}]{Istrate16}
{Istrate} A.~G.,  {Marchant} P.,  {Tauris} T.~M.,  {Langer} N.,  {Stancliffe}
  R.~J.,   {Grassitelli} L.,  2016, \mn@doi [\aap]
  {10.1051/0004-6361/201628874}, \href
  {https://ui.adsabs.harvard.edu/abs/2016A&A...595A..35I} {595, A35}

\bibitem[\protect\citeauthoryear{{Ivanova} et~al.,}{{Ivanova}
  et~al.}{2013}]{Ivanova13}
{Ivanova} N.,  et~al., 2013, \mn@doi [\aapr] {10.1007/s00159-013-0059-2}, \href
  {https://ui.adsabs.harvard.edu/abs/2013A&ARv..21...59I} {21, 59}

\bibitem[\protect\citeauthoryear{{Juri{\'c}} et~al.,}{{Juri{\'c}}
  et~al.}{2008}]{Juric2008}
{Juri{\'c}} M.,  et~al., 2008, \mn@doi [\apj] {10.1086/523619}, \href
  {https://ui.adsabs.harvard.edu/abs/2008ApJ...673..864J} {673, 864}

\bibitem[\protect\citeauthoryear{{Kordopatis} et~al.,}{{Kordopatis}
  et~al.}{2011}]{Kordo2011}
{Kordopatis} G.,  et~al., 2011, \mn@doi [\aap] {10.1051/0004-6361/201117373},
  \href {https://ui.adsabs.harvard.edu/abs/2011A&A...535A.107K} {535, A107}

\bibitem[\protect\citeauthoryear{Korol, Rossi, Groot, Nelemans, Toonen  \&
  Brown}{Korol et~al.}{2017}]{korol2017}
Korol V.,  Rossi E.~M.,  Groot P.~J.,  Nelemans G.,  Toonen S.,   Brown A.~G.,
  2017, Monthly Notices of the Royal Astronomical Society, 470, 1894

\bibitem[\protect\citeauthoryear{{Kov{\'a}cs}, {Zucker}  \&
  {Mazeh}}{{Kov{\'a}cs} et~al.}{2002}]{kovacs2002}
{Kov{\'a}cs} G.,  {Zucker} S.,   {Mazeh} T.,  2002, \mn@doi [\aap]
  {10.1051/0004-6361:20020802}, \href
  {https://ui.adsabs.harvard.edu/abs/2002A&A...391..369K} {391, 369}

\bibitem[\protect\citeauthoryear{{Kupfer} et~al.,}{{Kupfer}
  et~al.}{2018}]{kupfer2018}
{Kupfer} T.,  et~al., 2018, \mn@doi [\mnras] {10.1093/mnras/sty1545}, \href
  {https://ui.adsabs.harvard.edu/abs/2018MNRAS.480..302K} {480, 302}

\bibitem[\protect\citeauthoryear{{Littlefair}, {Dhillon}, {Marsh},
  {G{\"a}nsicke}, {Southworth}, {Baraffe}, {Watson}  \&
  {Copperwheat}}{{Littlefair} et~al.}{2008}]{Littlefair08}
{Littlefair} S.~P.,  {Dhillon} V.~S.,  {Marsh} T.~R.,  {G{\"a}nsicke} B.~T.,
  {Southworth} J.,  {Baraffe} I.,  {Watson} C.~A.,   {Copperwheat} C.,  2008,
  \mn@doi [\mnras] {10.1111/j.1365-2966.2008.13539.x}, \href
  {https://ui.adsabs.harvard.edu/abs/2008MNRAS.388.1582L} {388, 1582}

\bibitem[\protect\citeauthoryear{{Masci} et~al.,}{{Masci}
  et~al.}{2019}]{masci2019}
{Masci} F.~J.,  et~al., 2019, \mn@doi [\pasp] {10.1088/1538-3873/aae8ac}, \href
  {https://ui.adsabs.harvard.edu/abs/2019PASP..131a8003M} {131, 018003}

\bibitem[\protect\citeauthoryear{{McAllister} et~al.,}{{McAllister}
  et~al.}{2019}]{mcallister19}
{McAllister} M.,  et~al., 2019, \mn@doi [\mnras] {10.1093/mnras/stz976}, \href
  {https://ui.adsabs.harvard.edu/abs/2019MNRAS.486.5535M} {486, 5535}

\bibitem[\protect\citeauthoryear{{Patterson} et~al.,}{{Patterson}
  et~al.}{2018}]{Patterson18}
{Patterson} J.,  et~al., 2018, \mn@doi [\pasp] {10.1088/1538-3873/aaa7e7},
  \href {https://ui.adsabs.harvard.edu/abs/2018PASP..130f4202P} {130, 064202}

\bibitem[\protect\citeauthoryear{{Pelisoli} \& {Vos}}{{Pelisoli} \&
  {Vos}}{2019}]{Pelisoli19}
{Pelisoli} I.,  {Vos} J.,  2019, \mn@doi [\mnras] {10.1093/mnras/stz1876},
  \href {https://ui.adsabs.harvard.edu/abs/2019MNRAS.488.2892P} {488, 2892}

\bibitem[\protect\citeauthoryear{{Ramsay} et~al.,}{{Ramsay}
  et~al.}{2018}]{Ramsay18}
{Ramsay} G.,  et~al., 2018, \mn@doi [\aap] {10.1051/0004-6361/201834261}, \href
  {https://ui.adsabs.harvard.edu/abs/2018A&A...620A.141R} {620, A141}

\bibitem[\protect\citeauthoryear{{Rebassa-Mansergas}
  et~al.,}{{Rebassa-Mansergas} et~al.}{2021}]{RebassaMansergas21}
{Rebassa-Mansergas} A.,  et~al., 2021, \mn@doi [\mnras]
  {10.1093/mnras/stab1559}, \href
  {https://ui.adsabs.harvard.edu/abs/2021MNRAS.505.3165R} {505, 3165}

\bibitem[\protect\citeauthoryear{{Schreiber} \& {G{\"a}nsicke}}{{Schreiber} \&
  {G{\"a}nsicke}}{2003}]{SchreiberGaensicke03}
{Schreiber} M.~R.,  {G{\"a}nsicke} B.~T.,  2003, \mn@doi [\aap]
  {10.1051/0004-6361:20030801}, \href
  {https://ui.adsabs.harvard.edu/abs/2003A&A...406..305S} {406, 305}

\bibitem[\protect\citeauthoryear{{Sharma} et~al.,}{{Sharma}
  et~al.}{2020}]{Sharma2020}
{Sharma} S.,  et~al., 2020, arXiv e-prints, \href
  {https://ui.adsabs.harvard.edu/abs/2020arXiv200406556S} {p. arXiv:2004.06556}

\bibitem[\protect\citeauthoryear{{Southworth}, {Tappert}, {G{\"a}nsicke}  \&
  {Copperwheat}}{{Southworth} et~al.}{2015}]{Southworth15}
{Southworth} J.,  {Tappert} C.,  {G{\"a}nsicke} B.~T.,   {Copperwheat} C.~M.,
  2015, \mn@doi [\aap] {10.1051/0004-6361/201425060}, \href
  {https://ui.adsabs.harvard.edu/abs/2015A&A...573A..61S} {573, A61}

\bibitem[\protect\citeauthoryear{{Szkody} et~al.,}{{Szkody}
  et~al.}{2010}]{Szkody10}
{Szkody} P.,  et~al., 2010, \mn@doi [\apj] {10.1088/0004-637X/716/2/1531},
  \href {https://ui.adsabs.harvard.edu/abs/2010ApJ...716.1531S} {716, 1531}

\bibitem[\protect\citeauthoryear{{Toloza} et~al.,}{{Toloza}
  et~al.}{2019}]{Toloza19}
{Toloza} O.,  et~al., 2019, \baas, \href
  {https://ui.adsabs.harvard.edu/abs/2019BAAS...51c.168T} {51, 168}

\bibitem[\protect\citeauthoryear{{Uthas}, {Knigge}, {Long}, {Patterson}  \&
  {Thorstensen}}{{Uthas} et~al.}{2011}]{Uthas11}
{Uthas} H.,  {Knigge} C.,  {Long} K.~S.,  {Patterson} J.,   {Thorstensen} J.,
  2011, \mn@doi [\mnras] {10.1111/j.1745-3933.2011.01061.x}, \href
  {https://ui.adsabs.harvard.edu/abs/2011MNRAS.414L..85U} {414, L85}

\bibitem[\protect\citeauthoryear{{Wolz} et~al.,}{{Wolz} et~al.}{2018}]{Wolz18}
{Wolz} M.,  et~al., 2018, \mn@doi [Open Astronomy] {10.1515/astro-2018-0011},
  \href {https://ui.adsabs.harvard.edu/abs/2018OAst...27...80W} {27, 80}

\bibitem[\protect\citeauthoryear{{Zorotovic}, {Schreiber}, {Garc{\'\i}a-Berro},
  {Camacho}, {Torres}, {Rebassa-Mansergas}  \& {G{\"a}nsicke}}{{Zorotovic}
  et~al.}{2014}]{Zorotovic14}
{Zorotovic} M.,  {Schreiber} M.~R.,  {Garc{\'\i}a-Berro} E.,  {Camacho} J.,
  {Torres} S.,  {Rebassa-Mansergas} A.,   {G{\"a}nsicke} B.~T.,  2014, \mn@doi
  [\aap] {10.1051/0004-6361/201323039}, \href
  {https://ui.adsabs.harvard.edu/abs/2014A&A...568A..68Z} {568, A68}

\bibitem[\protect\citeauthoryear{{van Roestel} et~al.,}{{van Roestel}
  et~al.}{2021a}]{vanRoestel21_5amcvns}
{van Roestel} J.,  et~al., 2021a, \mn@doi [\mnras] {10.1093/mnras/stab2421},
  \href {https://ui.adsabs.harvard.edu/abs/2021MNRAS.tmp.2205V} {}

\bibitem[\protect\citeauthoryear{{van Roestel} et~al.,}{{van Roestel}
  et~al.}{2021b}]{vanRoestel21_WDBD}
{van Roestel} J.,  et~al., 2021b, arXiv e-prints, \href
  {https://ui.adsabs.harvard.edu/abs/2021arXiv210508687V} {p. arXiv:2105.08687}

\makeatother
\end{thebibliography}

\appendix
\section{Tables and Spectra}
\label{sec:appendix}

\begin{figure*}
    \includegraphics[width=0.90\linewidth]{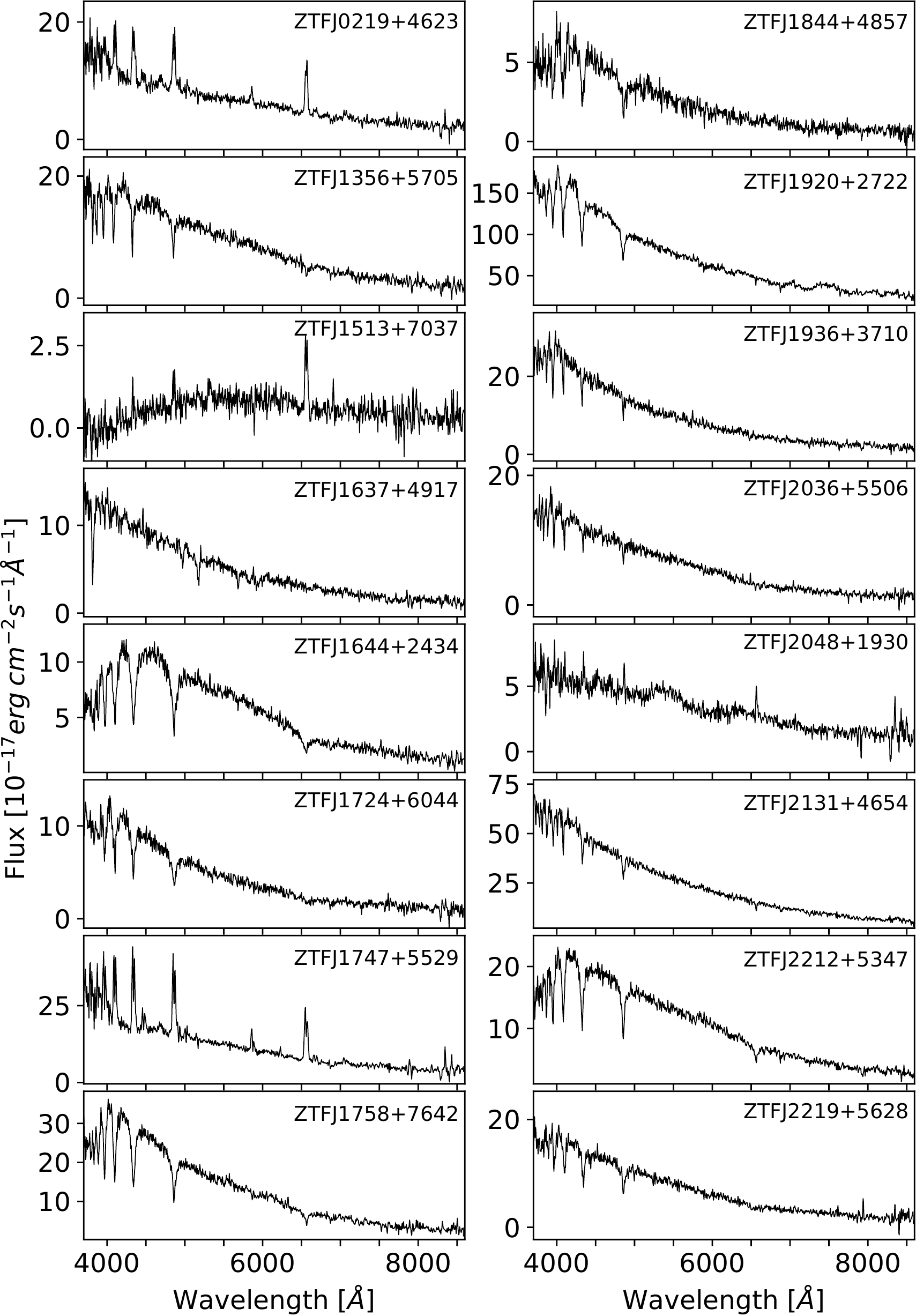}
	\caption{Identification spectra of 16 of the 18 newly discovered binaries, as observed with the Isaac Newton Telescope. The spectra are lightly smoothed with a Savitzky–Golay filter for display purposes. 
	\label{fig:spectra}}
\end{figure*}


\begin{table*}
\caption{Log of spectroscopic observations \label{tab:specobs} All spectra were obtained using the Intermediate Dispersion Spectrograph (IDS) on the Isaac Newton Telescope in La Palma, Spain. The time of the observation refers to the middle of the exposure.}
\begin{tabular*}{\textwidth}{lcrcll}
\hline
  \multicolumn{1}{c}{} &
  \multicolumn{1}{c}{MAGNITUDE} &
  \multicolumn{1}{c}{EXP.} &
  \multicolumn{1}{c}{OBSERVATION} &
  \multicolumn{1}{c}{WEATHER} &
  \multicolumn{1}{c}{} \\
  \multicolumn{1}{c}{ZTF Name} &
  \multicolumn{1}{c}{(Gaia G)} &
  \multicolumn{1}{c}{(s)} &
  \multicolumn{1}{c}{DATE (UT)} &
  \multicolumn{1}{c}{\& SEEING} &
  \multicolumn{1}{c}{COMMENT} \\
\hline
J0219+4623 & 19.77 & 3600 & 2021 Aug 6, 04:11:53 & Clear, $\sim0.9\arcsec$ & 
  Double-peaked H emission, single-peaked HeI. CV\\
J1356+5705 & 18.99 & 2400 & 2021 Aug 4, 21:53:07 & Clear, $\sim1.0\arcsec$ & 
  Balmer absorption, likely ELM\\
J1513+7037 & 19.48 & 3600 & 2021 Aug 6, 22:16:11 & Clear, $\sim2.0\arcsec$ & 
  Double-peaked H$\alpha$ and H$\beta$ emission. CV.\\
J1637+4917 & 19.40 & 2700 & 2021 Aug 4, 22:48:56 & Clear, $\sim1.1\arcsec$ & 
  Lines of HeI, MgI and NaI. AM\,CVn star\\
J1644+2434 & 19.13 & 2400 & 2021 Aug 6, 23:36:34 & Clear, $\sim0.9\arcsec$ & 
  Broad Balmer absorption. DA white dwarf. No companion features.\\
J1724+6044 & 20.11 & 3600 & 2021 Aug 5, 23:01:15 & Clear, $\sim0.7\arcsec$ & 
  Balmer absorption, weak/no H$\alpha$. DA white dwarf. \\
J1747+5529 & 19.25 & 2700 & 2021 Aug 6, 00:04:21 & Clear, $\sim0.8\arcsec$ & 
  Double-peaked H and HeI emission lines. CV\\
J1758+7642 & 19.00 & 2700 & 2021 Aug 5, 21:54:54 & Clear, $\sim1.0\arcsec$ & 
  Broad Balmer absorption. DA white dwarf. No companion features.\\
J1844+4857 & 19.33 & 2700 & 2021 Aug 4, 23:47:44 & Clear, $\sim1.6\arcsec$ & 
  Faint DA white dwarf spectrum. No  H$\alpha$ seen.\\
J1920+2722 & 15.58 &  600 & 2021 Aug 5, 00:30:42 & Clear, $\sim2.5\arcsec$ & 
  White dwarf--M dwarf composite spectrum.\\
J1936+3710 & 17.92 & 1600 & 2021 Aug 5, 02:13:02 & Clear, $\sim1.7\arcsec$ & 
  Hot subdwarf\\
J2036+5506 & 19.70 & 3600 & 2021 Aug 6, 02:01:13 & Clear, $\sim0.7\arcsec$ & 
  Weak and narrow Balmer lines.\\
J2048+1930 & 19.26 & 2700 & 2021 Aug 6, 01:00:40 & Clear, $\sim0.8\arcsec$ & 
  Balmer and HeI emission, broad cyclotron hump. Magnetic CV\\
J2131+4654 & 17.42 & 1200 & 2021 Aug 5, 02:54:56 & Clear, $\sim1.2\arcsec$ & 
  Hot subdwarf\\
J2212+5347 & 18.31 & 2400 & 2021 Aug 5, 03:32:06 & Clear, $\sim1.3\arcsec$ & 
  Balmer absorption, DA white dwarf. Companion not seen.\\
J2219+5628 & 19.66 & 2700 & 2021 Aug 6, 03:02:05 & Clear, $\sim0.8\arcsec$ & 
  Balmer absorption, no/weak H$\alpha$. \\
\hline
\end{tabular*}
\end{table*}


\begin{table*}
\caption{{\em GALEX} ultraviolet and PanSTARRS optical magnitudes of the 18 new binaries detected, in order of right ascension. The numbers in parenthesis indicate the uncertainty on the decimal digits.\label{tab:phot}}
\centering
\begin{tabular*}{0.7\textwidth}{lccccccc}
\hline
ZTF &  FUV &  NUV &  $g$ &  $r$ &  $i$ &  $z$ &  $y$ \\
\hline\hline
J0219+4623 &            &            &  19.94(03) &  19.75(04) &  19.64(02) &  19.57(13) &  19.59(14) \\
J0609+3652 &            &            &  18.74(01) &  18.91(03) &  19.26(03) &  19.39(05) &  19.32(06) \\
J1100+5210 &  17.64(07) &  17.92(04) &  18.38(02) &  18.72(01) &  18.98(01) &  19.13(05) &  19.25(09) \\
J1356+5705 &            &  20.29(11) &  19.08(02) &  19.00(01) &  19.06(01) &  19.15(03) &  19.20(05) \\
J1513+7037 &            &            &  19.81(04) &  19.26(02) &  19.41(09) &  19.39(05) &  19.19(02) \\
J1637+4917 &            &  20.34(09) &  19.33(01) &  19.52(02) &  19.60(01) &  19.75(03) &  19.83(05) \\
J1644+2434 &  20.01(16) &  19.70(08) &  19.00(01) &  19.21(01) &  19.38(01) &  19.48(03) &  19.30(04) \\
J1724+6044 &  19.52(12) &  19.76(09) &  19.95(02) &  20.24(02) &  20.38(03) &  20.37(05) &  19.97(07) \\
J1747+5529 &  21.50(29) &  20.77(16) &  19.10(03) &  19.08(03) &  19.11(02) &  18.82(00) &  18.80(07) \\
J1758+7642 &  19.27(14) &  19.38(10) &  18.88(00) &  19.11(01) &  19.33(01) &  19.50(02) &  19.74(23) \\
J1844+4857 &  19.43(12) &  19.52(06) &  19.29(01) &  19.48(02) &  19.51(02) &  19.30(04) &  19.12(05) \\
J1920+2722 &            &            &  15.67(01) &  15.85(03) &  15.68(02) &  15.34(00) &  15.29(02) \\
J1936+3710 &            &            &  17.76(01) &  18.08(01) &  18.37(01) &  18.45(03) &  18.61(03) \\
J2036+5506 &            &            &  19.72(02) &  19.78(02) &  19.99(03) &  20.04(04) &  20.14(27) \\
J2048+1930 &  21.34(34) &  20.78(17) &  19.25(09) &  19.23(06) &  19.30(06) &  19.33(08) &  19.31(08) \\
J2131+4654 &            &            &  17.43(01) &  17.48(01) &  17.69(02) &  17.76(02) &  17.98(02) \\
J2212+5347 &            &            &  18.39(01) &  18.38(00) &  18.39(01) &  18.29(01) &  18.09(05) \\
J2219+5628 &            &            &  19.68(02) &  19.74(02) &  19.88(02) &  19.92(05) &  19.61(04) \\
\hline
\end{tabular*}
\end{table*}


\begin{table*}
\caption{Properties of the nine known eclipsing binaries re-identified in this search. The $G$ band absolute magnitude was calculated using the distances derived by \citet{bj18}. References for the classifications are given in Section~\ref{sec:completeness}. \label{tab:knownbinaries}}
\renewcommand{\arraystretch}{1.3} 
\begin{tabular*}{\textwidth}{l| @{\extracolsep{\fill}} cccccccccccccccccc}
\hline\hline
\multicolumn{1}{c|}{Name} & Gaia DR2 & RA & DEC & $\omega$ & $G_\mathrm{BP}$--$G_\mathrm{RP}$ & $G_\mathrm{abs}$ & $P$ & $T_0$ & Class \\
~ & sourceID & (deg) & (deg) & (mas) & (mag) & (mag) & (days) & (BMJD(TDB)) & \\\hline
J0113+2257 & 2791084432881749760 & 18.41288 & 22.960843 & 0.70$\pm$0.09 & -0.38 & $6.01^{+0.30}_{-0.27}$ & 0.09337 & 58270.49979842 & sdB+M\\
J0538+1953 & 3402486457838073728 & 84.51134 & 19.884137 & 1.15$\pm$0.48 & -0.12 & $8.68^{+1.88}_{-1.16}$ & 0.010030 & 58734.23560684 & DWD\\
J0722-1839 & 2931430078489330944 & 110.58950 & -18.658474 & 0.41$\pm$0.35 & -0.01 & $7.10^{+1.40}_{-1.25}$ & 0.01646 & 58204.20323446 & DWD\\
J0935+1619 & 619153448780495872 & 143.90610 & 16.330769 & 0.51$\pm$0.34 & 0.19 & $8.19^{+0.79}_{-0.69}$ & 0.06405 & 58202.33255548 & CV \\
NZ Boo & 1289860214647954816 & 225.67043 & 33.573176 & 5.36$\pm$0.09 & 0.36 & $11.00^{+0.04}_{-0.03}$ & 0.05891 & 58198.38575829 & CV\\
OV Boo & 1593140224924964864 & 226.84215 & 52.511234 & 4.74$\pm$0.14 & 0.26 & $11.62^{+0.07}_{-0.06}$ & 0.04626 & 58198.38159296 & CV \\
Gaia14aae & 1629388752470472704 & 242.89152 & 63.142130 & 3.87$\pm$0.16 & 0.22 & $11.08^{+0.09}_{-0.09}$ & 0.03452 & 58198.39053330 & AM CVn\\
J1901+5309 & 2134541781964072320 & 285.35592 & 53.158151 & 1.11$\pm$0.12 & -0.33 & $8.33^{+0.25}_{-0.22}$ & 0.02820 & 58197.45747046 & DWD\\
WZ Sge & 1809844934461976832 & 301.90242 & 17.703988 & 22.16$\pm$0.04 & 0.15 & $11.94^{+0.00}_{-0.00}$ & 0.02834 & 58206.49033498 & CV\\
\hline
\end{tabular*}
\end{table*}


\begin{table*}
\caption{Eclipsing binaries detected by \citet{Burdge20} but not recovered in our search, mainly because they are not included in the \textit{Gaia} white dwarf catalogue. The ephemerides are from the spectroscopic orbits measured in that work, apart from J1905+3134, for which we calculated it from the ZTF data. An asterisk indicates an uncertain classification. } \label{tab:notinwdcat}
\renewcommand{\arraystretch}{1.3} 
\begin{tabular*}{\textwidth}{l| @{\extracolsep{\fill}} ccccccccccccccccc}
\hline\hline
\multicolumn{1}{c|}{ZTF Name} & Gaia DR2 & RA & DEC & $\omega$ & $G_\mathrm{BP}$--$G_\mathrm{RP}$ & $G_\mathrm{abs}$ & $P$ & $T_0$ & Class\\
~ & sourceID & (deg) & (deg) & (mas) & (mag) & (mag) & (days) & (BMJD(TDB)) &\\\hline
J1539+5027 & 1402814555998561280 & 234.88397 & 50.460757 & -0.11$\pm$0.79 & -0.09 & $9.78^{+1.06}_{-1.05}$ & 0.004801 & 58305.18278861 & DWD \\
J1749+0924 & 4488756396492138880 & 267.48037 & 9.409030 & -2.68$\pm$1.36 & 0.14 & $7.92^{+1.23}_{-1.40}$ & 0.018357 & 58634.410862 & DWD/ELM\\
J1905+3134 & 2042772624503621632 & 286.29723 & 31.575656 & -- & 0.02 & -- & 0.011946 & 58197.47200923 & AM\,CVn \\
J1946+3203 & 2034582328046714880 & 296.51620 & 32.053628 & 0.72$\pm$0.35 & 0.44 & $7.30^{+1.27}_{-1.44}$ & 0.023308 & 58836.3882513 & DWD$^*$ \\
J2029+1534 & 1805551543400732544 & 307.34292 & 15.575226 & -1.43$\pm$1.43 & 0.15 & $8.88^{+2.33}_{-1.53}$ & 0.014491 & 58757.25537855 & DWD \\
J2055+4651 & 2163801981105229184 & 313.81657 & 46.851772 & 0.43$\pm$0.10 & 0.32 & $5.97^{+0.55}_{-0.45}$ & 0.039130 & 58731.9444251 & WD-sdB\\
\hline
\end{tabular*}
\end{table*}

%


\bsp	
\label{lastpage}
\end{document}